\documentclass[runningheads]{llncs}

\usepackage{eccv}

\usepackage{eccvabbrv}
\usepackage{algorithm}
\usepackage{algpseudocode}
\usepackage{graphicx}
\usepackage{booktabs}
\usepackage{multirow} 
\usepackage[accsupp]{axessibility}  
\usepackage{subfloat}
\usepackage{subcaption}
\usepackage{caption}
\definecolor{darkblue}{rgb}{0.0, 0.0, 0.55}
\definecolor{cadmiumgreen}{rgb}{0.0, 0.42, 0.24}

\usepackage{hyperref}

\usepackage{orcidlink}

\begin{document}


\title{Stealthy Multi-task Adversarial Attacks} 

\titlerunning{Stealthy Multi-task Adversarial Attacks}

\author{Jiacheng Guo\inst{1,2*}\orcidlink{0009-0006-8508-9949} \and
Tianyun Zhang\inst{1*\dagger}\orcidlink{0000-0002-2475-6414} \and
Lei Li\inst{1}\orcidlink{0009-0003-4317-9071} \and
Haochen Yang\inst{1}\orcidlink{0000-0001-9145-8575} \and
Hongkai Yu\inst{1}\orcidlink{0000-0001-5383-8913} \and
Minghai Qin\inst{1,3\dagger}\orcidlink{0000-0001-5172-5309}}

\authorrunning{J.~Guo et al.}

\institute{Cleveland State University, Cleveland, OH 44115, USA \email{\{j.guo58,l.li15,h.yang15\}@vikes.csuohio.edu}, \email{\{t.zhang85,h.yu19\}@csuohio.edu} \and
University of Wisconsin-Madison, Madison, WI 53706, USA \and
Western Digital Research, San Jose, CA 95119, USA \\ \email{\{minghai.qin\}@wdc.com}\\
$^{*}$Equal Contribution $^{\dagger}$Corresponding Authors
}

\maketitle

\begin{abstract}
Deep neural networks are highly vulnerable to adversarial perturbations, raising serious safety concerns in the real-world systems. While prior work mainly explores single-task attacks or jointly degrading all tasks in multi-task models, practical scenarios often demand more selective and stealthy attack strategies. To address this challenge, we propose \textbf{S}tealthy \textbf{M}ulti-\textbf{T}ask \textbf{A}dversarial \textbf{A}ttack (\textbf{SMTA$^{2}$}), a novel framework that selectively degrades a targeted task while strictly preserving the performance of non-targeted tasks. We formulate this objective as a constrained multi-objective optimization problem and design task-aware adversarial perturbations that maximize degradation on the targeted task without causing collateral damage on non-targeted tasks. To enhance practicality, we further introduce an automated loss-weight tuning strategy that dynamically balances attack and preservation objectives. Experiments on two multi-task benchmarks NYUv2 and Cityscapes demonstrate that SMTA$^{2}$ achieves strong attack performance on targeted tasks while maintaining non-targeted tasks intact on both undefended and adversarially trained models, establishing the first systematic framework for stealthy and selective multi-task attack framework.

\end{abstract}

\section{Introduction}

\begin{figure*}[!ht]
\centering
\includegraphics[width=1\linewidth]{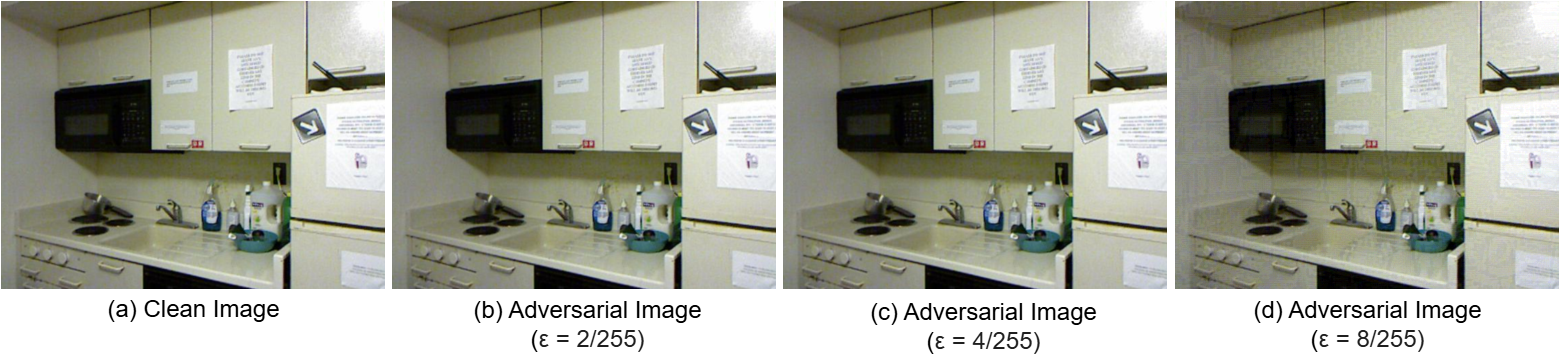} 
\caption{Demonstration of a clean image (a) and adversarial images (b) and (c) from NYUv2 dataset.}
\label{Atk_perturb}
\end{figure*}

Deep Neural Networks (DNNs) have achieved remarkable success across a wide range of tasks~\cite{lecun1998gradient,dahl2011context,hinton2012deep,andor2016globally}, yet remain highly vulnerable to adversarial attacks~\cite{madry2018towards,hu2026dynamic}. These attacks introduce subtle, often imperceptible perturbations that can drastically alter model predictions~\cite{nguyen2015deep,carlini2016hidden,goodfellow2014explaining}. Despite their minimal visual impact, such perturbations can cause severe performance degradation, posing significant safety risks in critical applications, such as autonomous driving~\cite{chakraborty2021survey,yang2025da3d}. Figure~\ref{Atk_perturb} illustrates typical imperceptible adversarial perturbations under different magnitudes, where even small perturbations (e.g., $\epsilon$=2/255 or 4/255) can effectively mislead DNNs.

Beyond single-task models, Multi-Task Learning (MTL) enables a shared network to jointly learn multiple related tasks, significantly improving model efficiency and generalization~\cite{caruana1997multitask,crawshaw2020multi,sener2018multi}. However, the shared representations also introduce coupled vulnerabilities: a perturbation crafted for one task can \textit{unintentionally propagate to others}~\cite{ghamizi2022adversarial,zhang2025attacking}. Specifically, in safety-critical systems, attackers rarely aim for indiscriminate performance collapse. Instead, they may selectively target high-value tasks while preserving auxiliary functionalities to evade detection~\cite{sun2024defense,zhe2024adversarial}. For instance, manipulating traffic sign recognition while preserving lane detection may create subtle yet dangerous misbehavior in autonomous driving~\cite{long2025physical}. Similarly, degrading pedestrian segmentation while keeping depth estimation intact can cause catastrophic consequences without triggering obvious system-wide failure~\cite{liu2023deep}. These scenarios highlight the need for \textit{selective} and \textit{stealthy} multi-task adversarial attacks that compromise targeted tasks while preserving the performance of others.

However, achieving \textit{selective degradation} is fundamentally challenging~\cite{zhang2025attacking,zhe2024adversarial}. Multi-task objectives are inherently coupled through shared parameters, and increasing the loss of one task typically propagates gradients that harm others. Enforcing strict preservation constraints while maximizing degradation on a targeted task leads to a highly constrained and non-trivial optimization problem. Naively reweighting losses often fails to maintain non-targeted task performance, especially under strong adversarial perturbations. 

To address these challenges, we propose a novel concept termed \textbf{Stealthy Multi-Task Adversarial Attack (SMTA$^{2}$)}. Given an MTL framework with multiple tasks, SMTA$^{2}$ aims to significantly degrade a \textbf{targeted task} while strictly preserving or even improving the performance of other tasks which are not expected to be attacked, regarded as \textbf{non-targeted tasks}. By crafting task-aware adversarial perturbations through a weighted loss formulation, SMTA$^{2}$ enables fine-grained control over task-specific attack behaviors. Furthermore, we develop an automated weight searching approach to dynamically search for optimal task weights, enabling efficient and adaptive attack generation without manual tuning. This strategy significantly reduces computational overhead and improves generalization across diverse inputs. Compared to random system noise which typically causes uniform and mild degradation across all tasks, SMTA$^{2}$ introduces asymmetric and task-selective degradation, a property that is empirically validated in Section~\ref{Experiments}. 

To the best of our knowledge, this is the first work that systematically formulates and solves the problem of stealthy and task-selective adversarial attacks in MTL, together with a principled evaluation criterion for preservation-aware attack assessment. The main contributions are summarized as follows:
\begin{itemize}
\item We propose ``Stealthy Multi-Task Adversarial Attacks'' (\textbf{SMTA$^{2}$}), a novel attack paradigm which enables selective degradation of a targeted task while strictly preserving the original performance of non-targeted tasks in MTL.

\item The ``Stealthy Multi-Task Adversarial Attacks'' is formulated as a constrained multi-objective optimization problem and designed as a task-aware objective with explicit preservation constraints.

\item We develop an automated loss-weight optimization strategy that dynamically searches for optimal task weights, improving attack efficiency and stability over manual tuning.

\item Experiments on NYUv2 and Cityscapes datasets in different $L_p$ norm attacks demonstrate strong targeted attack effectiveness while preserving non-targeted tasks on both undefended and adversarially trained models.

\end{itemize}

\section{Related Work}

\subsection{Multi-task Learning}

Multi-task learning (MTL) is a machine learning paradigm which enables the simultaneous training of multiple related tasks, providing advantages such as improved data efficiency, reduced overfitting, and enhanced generalization capabilities \cite{caruana1997multitask,crawshaw2020multi}. By leveraging shared representations among tasks, MTL has become increasingly popular across diverse fields, including natural language processing and computer vision \cite{zhang2023survey,yu2024unleashing,jia2026adaptive}. Recent advancements have categorized MTL techniques into several domains, such as regularization strategies, task relationship modeling, feature propagation, optimization frameworks, and pre-training methods \cite{hinton2012deep,guo2024min,guo2025min}. 

MTL approaches are broadly classified into joint training and multi-step training methods, depending on the nature of task relationships and the level of interaction between tasks \cite{zhang2023survey,li2025task}, which has evolved significantly with the emergence of deep learning and pretrained foundation models. These advancements have introduced innovations such as task-promotable, task-agnostic training, and capabilities for zero-shot and few-shot learning, making MTL applicable in scenarios with limited task-specific labeled data \cite{he2020contrastive,yu2024unleashing}. As MTL continues to mature, emerging research focuses on improving task balancing, addressing task interference, and enhancing robustness for its up-to-date application in complex domains~\cite{li2025task,li2026fedsta}.

\subsection{Adversarial Attacks}

Adversarial attacks pose a fundamental challenge to DNNs, where carefully crafted perturbations can induce incorrect predictions despite being largely imperceptible to humans~\cite{sun2024defense,xu2026etv,nguyen2015deep,carlini2017towards,madry2018towards}. Such perturbations expose critical vulnerabilities in modern architectures.

Gradient-based optimization remains a dominant paradigm for attack generation. Universal adversarial perturbations~\cite{moosavi2017universal} demonstrated that input-agnostic, gradient-derived perturbations can reliably fool classifiers, laying the foundation for subsequent gradient-driven methods and extensions beyond classification. These approaches enabled adversarial attacks to be studied in structured and real-world settings, including aerial object detection~\cite{du2022physical,xu2024adversarial}, semantically stealthy segmentation~\cite{chen2022semantically}, and audio tasks~\cite{jia2026adaptive} where perturbations preserve visual plausibility while inducing substantial degradation. In MTL, Guo et al.~\cite{guo2020multi} introduced a Multi-Task Adversarial Attack (MTA) framework that exploits shared representations to craft joint attacks, outperforming single-task baselines. Zhang et al.~\cite{zhang2025attacking} later standardized MTA methodologies and broadened evaluations. Nevertheless, unlike those non-stealthy architectures~\cite{guo2020multi,zhang2025attacking}, our objective is stricter: selectively attacking a targeted task and strictly preserving non-targeted tasks.

\subsection{Adversarial Robustness in MTL}

To improve adversarial robustness, Madry et al.~\cite{madry2018towards} formulated defense as a min-max optimization problem, while Mao et al.~\cite{mao2021adversarial} introduced natural supervision to recover correct predictions under attack.

Adversarial vulnerabilities in MTL have also been investigated. Zhang et al.~\cite{zhang2019theoretically} analyzed the robustness–accuracy trade-off theoretically. Zhang et al.~\cite{zhang2025attacking} proposed Gradient Balancing Multi-Task Attack to jointly optimize attacks across tasks, improving task balancing through shared representations. Guo et al.~\cite{guo2025robust} incorporated min–max optimization to enhance robustness in multi-task adversarial learning. These methods mainly focus on jointly attacking all tasks through improved optimization. Moreover, Zhe et al.~\cite{zhe2024adversarial} predefined ``hidden tasks'' which can be stealthily compromised via shared representations without explicitly modeling task relationships, but they used different task definitions and settings. Comparatively, we propose a unique SMTA$^2$ framework which attacks multi-task models without changing architectures or task relationships. 

Overall, prior works either design general optimization-based attacks or indiscriminately degrade all tasks, leaving selective and stealthy mechanisms underexplored. Crucially, the structural dependency induced by shared representations has not been systematically exploited for controlled adversarial manipulation. We address this gap by proposing a principled framework for selective and stealthy adversarial attacks in multi-task scenarios that explicitly models inter-task relationships, enabling fine-grained degradation while preserving all non-targeted tasks.

\section{Methodology}

\subsection{Problem Formulation}

Consider a multi-task learning setting where each input sample $x$ is associated with multiple task-specific ground-truth labels $y = (y_{1}, y_{2}, ..., y_{m})$. Let $L_i(x, y_i)$ denote the loss corresponding to the $i$-th task in a pretrained multi-task model with shared representations. 

\textbf{Definition.} The \textbf{Stealthy Multi-Task Adversarial Attack (SMTA$^{2}$)} is a kind of adversarial attack which aims to degrade the performance of a designated \emph{targeted task} while strictly preserving the performance of all remaining \emph{non-targeted tasks}.

Let $i_t$ be the targeted task. The SMTA$^{2}$ can be formulated as 
\begin{equation}
\label{orig}
\begin{aligned}
& \underset{ \delta }{\text{maximize}}
& & L_{i_t}(  x+ \delta,  y_{i_t} ),
\\ & \text{subject to}
& & L_{i}(  x+ \delta,  y_{i} ) \leq L_i(  x,  y_i )~\text{for}~i \neq i_t,
\\ & & & ||\delta||_p \leq \epsilon, x+\delta \in \mathcal B,
\end{aligned}
\end{equation}
where $\delta$ is the adversarial noise, $\epsilon$ is a value to constrain the strength of the adversarial noise, and $\mathcal B$ is the box constraint to ensure that a valid image is derived after adding the adversarial noise. 

The attack without the first constraint, $L_{i}(  x+ \delta,  y_{i} ) \leq L_i(  x,  y_i )~\text{for}~i \neq i_t,$ is to be called a \textbf{non-stealthy} multi-task attack.

\paragraph{Optimization Challenge.}
Problem~(\ref{orig}) is substantially more challenging than standard adversarial attacks. In multi-task DNNs, tasks share a common backbone representation, leading to strong inter-task coupling. Perturbations that increase the loss of one task naturally propagate to others. Therefore, the feasible region satisfying all preservation constraints is often narrow under strong perturbations, making direct optimization intractable. 

To obtain a tractable surrogate, we relax problem (\ref{orig}) to the following weighted objective:
\begin{equation}
\label{relax}
  \underset{ ||\delta||_p \leq \epsilon, x+\delta \in \mathcal B }{\text{maximize}} ~ ~ L (x+\delta, y) = \sum_{i=1}^m w_i L_i(  x+\delta,  y_i ),
\end{equation}
where $w_i$ is the weighting factor corresponding to the $i$-th task. If $i = i_t$, $w_i$ is positive; otherwise $w_i$ is negative. The constraint in Eq. (\ref{orig}) serves as an effective optimization proxy so that an appropriate combination of weighting factors corresponding to different tasks could be searched each time for metric preservation when solving problem (\ref{relax}). It is guaranteed that an optimal solution feasible to problem (\ref{orig}) could be found out by adjusting the weighting factors. Specifically, when the loss of a non-targeted task increases, its corresponding negative weight effectively penalizes further degradation. Crucially, success hinges on carefully adapting $\{w_i\}$ to balance attack strength and strict preservation, which is an inherently non-trivial task due to inter-task gradient interference.

Unlike conventional adversarial attacks which optimize only a single objective, SMTA$^{2}$ must simultaneously maximize one task loss while enforcing multiple preservation constraints. Due to the shared representations and gradient interference across tasks, satisfying these constraints while maintaining strong attack effectiveness is fundamentally challenging. In this case, it is crucial to find an optimal combination to balance different tasks. Therefore, effective weight adaptation strategies are essential for navigating this coupled optimization landscape. The details of how to satisfy the constraints in the stealthy attack is discussed in the subsection~\ref{SMTA_framework}.

\subsection{The Stealthy Multi-Task Adversarial Attack Framework}
\label{SMTA_framework}

In a general multi-task adversarial attack framework, attacking one task could negatively affect all other tasks facing severe degradation as well. As a result, the performance of non-targeted tasks could also be significantly affected. To maintain the performance of non-targeted tasks, a positive weighting factor is set for the targeted task and negative weighting factors for the non-targeted tasks when solving problem (\ref{relax}). The positive weighting factor can degrade the performance of the targeted task, while the negative ones could compensate for the performance degradation of non-targeted tasks caused by attacking the targeted task. If the magnitudes of negative weighting factors are too small, the performance of non-targeted tasks will still be degraded. In this case, the first constraint in problem (\ref{orig}) is not satisfied. If the negative weights are too large, the performance of non‑targeted tasks may even surpass the clean baseline, satisfying the preservation constraint but weakening the attack on the targeted task. and the solution of problem (\ref{relax}) will satisfy the constraints of problem (\ref{orig}). However, the strength of attacking the targeted task will be compromised. 

In SMTA$^{2}$, the core challenge lies in enforcing strict performance constraints on non-targeted tasks while maximizing the degradation of the targeted task. This naturally leads to a constrained multi-objective optimization problem, where the loss of the targeted task is maximized subject to preservation constraints on the remaining tasks. To address this challenge, we design two strategies for searching appropriate task weights in the joint loss function: a \textit{manual} search strategy and a more efficient \textit{automated} optimization strategy.

\paragraph{Manual Weight Searching.}
In the manual approach, the weight of the targeted task is fixed to emphasize its degradation, while the weights of non-targeted tasks are manually adjusted to enforce preservation constraints. The procedure begins by assigning small-magnitude weights to non-targeted tasks. After performing the adversarial attack, if the performance of any non-targeted task violates the preservation constraint (i.e., performance drops below the baseline), its corresponding weight magnitude is increased in the next trial. This iterative process continues until all non-targeted tasks satisfy the constraint.

Although this approach guarantees the existence of a feasible solution to Problem~(\ref{orig}), it relies heavily on trial-and-error hyperparameter tuning. Particularly, as the number of tasks increases, the search space grows rapidly, making manual tuning computationally expensive and time-consuming.

\begin{algorithm}[!ht]
\caption{An automated approach for searching the weighting factors in the loss function}
\begin{algorithmic}
\State Input: input data $x$, the ground-truth label of the i-th task $y_i$, the initialized weighting factor $w_i$ of the $i$-th task, model parameters and the corresponding loss $L (x, y) =  \sum_{i=1}^m w_i L_i(  x,  y_i )$, total steps $S$ in the adversarial attack, the targeted task $i_t$, the step size $\lambda$ to adjust weighting factors, and the attenuation coefficient $\alpha$. 
\State Initialize $\lambda_i=\lambda$ for all $i \neq i_t$.
\For{ $s =  1,2,\ldots, S$}
\State perform one step PGD (or APGD or IFGSM) attack on loss $L(x, y)$ to generate the updated adversarial noise $\delta$.
\For{$i =  1,2,\ldots, m$}
\If{$i \neq i_t$}
\State $w_i = w_i-\lambda_i*(L_i(  x + \delta,  y_i )-L_i(  x,  y_i ))$.

\If{$L_i(  x + \delta,  y_i ) \leq L_i(  x,  y_i )$}
\State {$\lambda_i = \lambda_i * \alpha$}

\EndIf

\EndIf
\EndFor
\EndFor

\end{algorithmic}\label{auto}
\end{algorithm}

\paragraph{Automated Weight Optimization.}
To overcome the inefficiency of manual tuning, we propose an automated weight optimization strategy, summarized in Algorithm~\ref{auto}. Unlike the manual approach, the automated weight searching approach dynamically updates task weights during each attack iteration, enabling real-time enforcement of preservation constraints. Specifically, at each attack step, we compute the loss difference between the adversarial and original inputs for each non-targeted task. The weight of each non-targeted task is updated proportionally to the loss difference. If adversarial loss increases (indicating potential degradation), the corresponding weight is adjusted to counteract this effect.

To accelerate convergence, we introduce a task-specific step size $\lambda_i$, initialized uniformly. Once a non-targeted task satisfies the preservation constraint (i.e., its adversarial loss no longer exceeds the original loss), its step size is attenuated by a coefficient $\alpha \in$(0,1), reducing oscillations and stabilizing the search process. This adaptive attenuation mechanism allows the algorithm to quickly converge to a feasible region while maintaining stability.

\paragraph{Complexity Analysis.}

Let $m$ denote the number of tasks and $S$ denote the total attack iterations. Each step of the attack requires one forward and backward pass to update the perturbation, incurring the same dominant cost attacks based on Projected Gradient Descent (PGD)~\cite{madry2018towards}, Auto-PGD~\cite{croce2021mind} or Iterative Fast Gradient Sign Method (IFGSM)~\cite{kurakin2018adversarial}. The additional overhead of SMTA$^{2}$ comes from updating task weights, which involves computing loss differences for ($m$-1) non-targeted tasks and simple scalar updates. This adds only $\mathcal{O}(m)$ operations per iteration, which is negligible compared to the backpropagation cost. Therefore, the overall time complexity remains $\mathcal{O}(S\cdot\text{Backprop})$, making SMTA$^{2}$ asymptotically equivalent to conventional attacks in complexity. Compared with manual weight searching strategy, the automated approach substantially improves efficiency and scalability, especially when the number of tasks increases. 

Overall, the automated approach performs online constrained optimization over task weights, significantly reducing parameter tuning time and improving scalability to multi-task settings with larger task counts. Empirically, it consistently finds feasible and stable solutions that satisfy stealthy attack requirements while preserving the performance among the non-targeted tasks.

\section{Experiments}
\label{Experiments}

To evaluate SMTA$^{2}$, we consider two schemes for searching optimal task weights: the \textit{manual} scheme and the \textit{automated} scheme. We adopt three widely-used attack algorithms: Projected Gradient Descent (PGD)~\cite{madry2018towards}, Iterative Fast Gradient Sign Method (IFGSM)~\cite{kurakin2018adversarial}, and a recently established stronger attack: Auto-PGD (APGD)~\cite{croce2021mind}. For the attack norms, the PGD attack is implemented under both $L_2$ and $L_\infty$ norms, the APGD is implemented under both $L_1$ and $L_2$ norms, and the IFGSM is under $L_\infty$ norm. We compare manual and automated weight-search scheme under the proposed loss formulation, respectively. Due to space limitation, the results of IFGSM and more visualization results are provided in the supplementary material.

\subsection{Experimental Setup}

We evaluate SMTA$^{2}$ on two widely used multi-task image benchmarks, i.e., NYUv2~\cite{silberman2012indoor} and Cityscapes~\cite{cordts2016cityscapes}. Both non-stealthy and stealthy attack settings are considered. Non-stealthy attacks degrade the targeted task and inevitably affects other tasks, causing global performance drops~\cite{guo2020multi,zhang2025attacking}. However, SMTA$^{2}$ aims to significantly impair the targeted task while strictly maintaining non-targeted task performance. We further incorporate adversarial training to assess robustness against defended models. All the images are resized to 288$\times$384 for NYUv2 and 256$\times$512 for Cityscapes. All the experiments are conducted on 8 NVIDIA RTX A6000 GPUs.

\subsection{Experimental Procedure}

The proposed SMTA$^{2}$ aims to selectively degrade a targeted task while strictly preserving non-targeted tasks. Accordingly, we adopt an evaluation criterion that measures attack effectiveness under full preservation constraints.

The attack strength is controlled via perturbations according to different $L_p$ norms. Specifically, for the $L_\infty$ norm in PGD and IFGSM, $\epsilon$ is set to 2/255, 4/255, and 8/255, respectively. The perturbation $\epsilon$ is set to 10 and 20 for $L_1$ norm in APGD and is set to 5 and 10 for $L_2$ norm in APGD and PGD. The manual strategy searches for feasible task weights, where negative weights for non-targeted tasks help enforce preservation. The automated strategy (Algorithm~\ref{auto}) could dynamically update task weights for efficient and stable convergence. Furthermore, we evaluate SMTA$^{2}$ on Adversarially Trained (AT) models. Results show that both strategies maintain non-targeted task performance while achieving strong targeted attacks, even under enhanced robustness. 

\paragraph{Evaluation Metrics.} On NYUv2, we evaluate semantic segmentation, depth estimation, and surface normal prediction tasks using mIoU, absolute error (aErr.), and mean angular distance (mDist.), respectively. On Cityscapes, mIoU is used for semantic segmentation and part segmentation tasks, and aErr. is used for disparity estimation. All the settings regarding metrics follow~\cite{liu2022auto}. Each task is treated as the targeted task in turn, with others considered as non-targeted.

\begin{table*}[!t]
\caption{Experimental results of non-stealthy and SMTA$^{2}$ framework by manual and automated solutions of \textbf{PGD $L_\infty$} attack algorithm on \textbf{NYUv2} dataset. Non-attacked performance is represented by $\epsilon$=0. The performances of the targeted task are underlined. Values affected by the attack (i.e., degraded during the attack) are marked in \textcolor{blue}{blue}, while those not influenced by the attack are highlighted in \textcolor{red}{red}. (↑) means higher is better and (↓) means lower is better.}
\centering
\resizebox{\linewidth}{!}{
\begin{tabular}{c|ccc|ccc|ccc}
\toprule
\begin{tabular}
{@{}c@{}} Method \end{tabular} & \multicolumn{3}{c|}{Non-stealthy} & \multicolumn{3}{c|}{SMTA$^{2}$ Manual} & \multicolumn{3}{c}{SMTA$^{2}$ Auto} \\ \midrule
\begin{tabular}
{@{}c@{}} Tasks \end{tabular} & \begin{tabular}[c]{@{}c@{}}Segment\\ \begin{small}{[}mIoU(↑){]}\end{small}\end{tabular} & \begin{tabular}[c]{@{}c@{}}Depth\\ \begin{small}{[}aErr(↓){]}\end{small}\end{tabular} & \begin{tabular}[c]{@{}c@{}}Normal\\ \begin{small}{[}mDist(↓){]}\end{small}\end{tabular} & \begin{tabular}[c]{@{}c@{}}Segment\\ \begin{small}{[}mIoU(↑){]}\end{small}\end{tabular} & \begin{tabular}[c]{@{}c@{}}Depth\\ \begin{small}{[}aErr(↓){]}\end{small}\end{tabular} & \begin{tabular}[c]{@{}c@{}}Normal\\ \begin{small}{[}mDist(↓){]}\end{small}\end{tabular} & \begin{tabular}[c]{@{}c@{}}Segment\\ \begin{small}{[}mIoU(↑){]}\end{small}\end{tabular} & \begin{tabular}[c]{@{}c@{}}Depth\\ \begin{small}{[}aErr(↓){]}\end{small}\end{tabular} & \begin{tabular}[c]{@{}c@{}}Normal\\ \begin{small}{[}mDist(↓){]}\end{small}\end{tabular} \\ \midrule
$\epsilon$=0 & 46.56  & 40.57                           & 23.41        & 46.56  & 40.57                           & 23.41   & 46.56  & 40.57                           & 23.41   \\ \midrule
& \textcolor{blue}{\underline{26.21}}                        & \textcolor{blue}{51.78}                &  \textcolor{blue}{26.65} & \textcolor{blue}{\underline{26.54}}   & \textcolor{red}{37.92}                        & \textcolor{red}{23.00} & \textcolor{blue}{\underline{26.46}}   & \textcolor{red}{40.28}                        & \textcolor{red}{23.12}         \\
$\epsilon$=2/255    &  \textcolor{blue}{35.08}   & \textcolor{blue}{\underline{106.93}}  &  \textcolor{blue}{28.45}  & \textcolor{red}{47.53}   &\textcolor{blue}{\underline{106.47}}   & 
 \textcolor{red}{23.19}  &   \textcolor{red}{47.63}   &\textcolor{blue}{\underline{106.45}}   & 
 \textcolor{red}{23.40}  \\  &  \textcolor{blue}{37.76}   &  \textcolor{blue}{54.99}   &  \textcolor{blue}{\underline{40.36}}   &  \textcolor{red}{48.17}   &  \textcolor{red}{38.23}   &   \textcolor{blue}{\underline{41.10}}  &  \textcolor{red}{46.76}   &  \textcolor{red}{34.76}   &   \textcolor{blue}{\underline{39.99}}   \\ 
 \midrule
& \textcolor{blue}{\underline{17.04}} 
 & \textcolor{blue}{63.80}  &  \textcolor{blue}{30.97}    &  \textcolor{blue}{\underline{18.30}}  &  \textcolor{red}{37.30}  &  \textcolor{red}{23.22}  &  \textcolor{blue}{\underline{18.33}}  &  \textcolor{red}{37.45}  &  \textcolor{red}{23.31}    \\ $\epsilon$=4/255   & \textcolor{blue}{24.15} 
  &  \textcolor{blue}{\underline{167.40}} 
  &   \textcolor{blue}{34.39}     & \textcolor{red}{47.08} 
 &  \textcolor{blue}{\underline{165.27}}  
&  \textcolor{red}{22.83}     & \textcolor{red}{47.67}
&  \textcolor{blue}{\underline{168.69}}  
&  \textcolor{red}{23.36}      \\           &  \textcolor{blue}{28.88}  &  \textcolor{blue}{71.44}  & \textcolor{blue}{\underline{53.88}}   & \textcolor{red}{47.86}   &   \textcolor{red}{37.26}   &   \textcolor{blue}{\underline{53.47}}  & \textcolor{red}{46.98}   &   \textcolor{red}{39.58}   &   \textcolor{blue}{\underline{53.87}}    \\ \midrule
&  \textcolor{blue}{\underline{10.11}}  &  \textcolor{blue}{78.36}  &  \textcolor{blue}{36.23}  &  \textcolor{blue}{\underline{12.96}}  &   \textcolor{red}{37.86}   &  \textcolor{red}{22.60}  &  \textcolor{blue}{\underline{12.32}}  &   \textcolor{red}{39.56}   &  \textcolor{red}{23.21} \\  $\epsilon$=8/255  & \textcolor{blue}{14.58} 
&  \textcolor{blue}{\underline{240.52}}   &  \textcolor{blue}{41.16}     &    \textcolor{red}{47.14} 
&  \textcolor{blue}{\underline{229.64}}  & \textcolor{red}{21.96}   &    \textcolor{red}{52.83} 
&  \textcolor{blue}{\underline{225.11}}  & \textcolor{red}{22.36} \\   & \textcolor{blue}{18.91}  &   \textcolor{blue}{93.36}   &   \textcolor{blue}{\underline{68.05}}  &   \textcolor{red}{47.66}  &  \textcolor{red}{39.93}  & \textcolor{blue}{\underline{66.60}} &   \textcolor{red}{46.93}  &  \textcolor{red}{38.03}  & \textcolor{blue}{\underline{63.29}}     \\
\bottomrule
\end{tabular}}
\label{Table_SMTA$^{2}$_pgdli_nyuv2}
\end{table*}

\begin{table*}[ht]
\caption{Results of non-stealthy attacks and SMTA$^{2}$ on \textbf{Cityscapes} dataset under \textbf{PGD $L_\infty$} attack. The targeted-task results are underlined. \textcolor{blue}{Blue} and \textcolor{red}{red} indicate affected and preserved values, respectively. (↑)/(↓) denote higher/lower is better.}
\centering
\resizebox{\linewidth}{!}{
\begin{tabular}{c|ccc|ccc|ccc}
\toprule
\begin{tabular}
{@{}c@{}} Method \end{tabular} & \multicolumn{3}{c|}{Non-stealthy} & \multicolumn{3}{c|}{SMTA$^{2}$ Manual} & \multicolumn{3}{c}{SMTA$^{2}$ Auto}\\ \midrule
\begin{tabular}
{@{}c@{}} Tasks \end{tabular} & \begin{tabular}[c]{@{}c@{}}Segment\\ \begin{small}{[}mIoU(↑){]}\end{small}\end{tabular} & \begin{tabular}[c]{@{}c@{}}Part Seg\\ \begin{small}{[}mIoU(↑){]}\end{small}\end{tabular} & \begin{tabular}[c]{@{}c@{}}Disparity\\ \begin{small}{[}aErr(↓){]}\end{small}\end{tabular} &
\begin{tabular}[c]{@{}c@{}}Segment\\ \begin{small}{[}mIoU(↑){]}\end{small}\end{tabular} & \begin{tabular}[c]{@{}c@{}}Part Seg\\ \begin{small}{[}mIoU(↑){]}\end{small}\end{tabular} & \begin{tabular}[c]{@{}c@{}}Disparity\\ \begin{small}{[}aErr(↓){]}\end{small}\end{tabular} & \begin{tabular}[c]{@{}c@{}}Segment\\ \begin{small}{[}mIoU(↑){]}\end{small}\end{tabular} & \begin{tabular}[c]{@{}c@{}}Part Seg\\ \begin{small}{[}mIoU(↑){]}\end{small}\end{tabular} & \begin{tabular}[c]{@{}c@{}}Disparity\\ \begin{small}{[}aErr(↓){]}\end{small}\end{tabular} \\ \midrule
$\epsilon$=0  & 54.20   &   51.82                         & 81.51  & 54.20   &   51.82                         & 81.51    & 54.20   &   51.82                         & 81.51 \\ \midrule
 & \textcolor{blue}{{\underline{27.98}}}                        & \textcolor{blue}{43.82}                &  \textcolor{blue}{98.72}  & \textcolor{blue}{\underline{25.34}}   & \textcolor{red}{52.51}                        & \textcolor{red}{79.71} & \textcolor{blue}{\underline{25.38}}   & \textcolor{red}{52.97}                        & \textcolor{red}{80.30}        \\
$\epsilon$=2/255   &  \textcolor{blue}{41.24}   & \textcolor{blue}{\underline{30.00}}  &  \textcolor{blue}{92.13}    & \textcolor{red}{54.57}   &\textcolor{blue}{\underline{27.29}}   &  \textcolor{red}{80.04} &   \textcolor{red}{54.25} &\textcolor{blue}{\underline{27.22}}   & 
\textcolor{red}{76.25} \\  &  \textcolor{blue}{38.47}   &  \textcolor{blue}{42.05}   &  \textcolor{blue}{\underline{278.11}}  & 
\textcolor{red}{54.39}   &  \textcolor{red}{51.93}   &   \textcolor{blue}{\underline{367.34}}  &  \textcolor{red}{56.16}   &  \textcolor{red}{52.30}   &   \textcolor{blue}{\underline{359.65}}    \\ 
 \midrule
 & \textcolor{blue}{\underline{20.48}} 
 & \textcolor{blue}{35.01}  &  \textcolor{blue}{119.67}     &  \textcolor{blue}{\underline{17.25}}  &  \textcolor{red}{52.67}  &  \textcolor{red}{81.00}  &  \textcolor{blue}{\underline{17.08}}  &  \textcolor{red}{52.77}  &  \textcolor{red}{81.21}   \\ $\epsilon$=4/255 & \textcolor{blue}{33.98} 
  &  \textcolor{blue}{\underline{21.69}} 
  &   \textcolor{blue}{111.19}       & \textcolor{red}{54.59} 
 &  \textcolor{blue}{\underline{18.02}}  
&  \textcolor{red}{80.23}     & \textcolor{red}{54.36}
 &  \textcolor{blue}{\underline{18.19}}  
 &  \textcolor{red}{80.00}    \\           &  \textcolor{blue}{25.82}  &  \textcolor{blue}{26.68}  & \textcolor{blue}{\underline{490.15}}     & \textcolor{red}{55.22}   &   \textcolor{red}{51.85}   &   \textcolor{blue}{\underline{679.89}}  & \textcolor{red}{56.44}   &   \textcolor{red}{52.02}   &   \textcolor{blue}{\underline{644.12}}  \\ \midrule
&  \textcolor{blue}{\underline{11.31}}  &  \textcolor{blue}{19.24}  &  \textcolor{blue}{164.19}  & 
 \textcolor{blue}{\underline{10.62}}  &   \textcolor{red}{51.84}   &  \textcolor{red}{78.72}  &  \textcolor{blue}{\underline{10.02}}  &   \textcolor{red}{52.07}   &  \textcolor{red}{81.50} \\  $\epsilon$=8/255  & \textcolor{blue}{25.77} 
&  \textcolor{blue}{\underline{14.65}}   &  \textcolor{blue}{153.10}    &    \textcolor{red}{55.13} 
&  \textcolor{blue}{\underline{12.63}}  & \textcolor{red}{79.12}   &    \textcolor{red}{54.27}
&  \textcolor{blue}{\underline{12.22}}  & \textcolor{red}{80.97}  \\   & \textcolor{blue}{16.05}  &   \textcolor{blue}{14.18}   &   \textcolor{blue}{\underline{824.75}}     &   \textcolor{red}{55.10}  &  \textcolor{red}{52.54}  & \textcolor{blue}{\underline{1014.74}}  &   \textcolor{red}{57.36}  &  \textcolor{red}{52.10}  & \textcolor{blue}{\underline{1008.44}} \\
\bottomrule
\end{tabular}}
\label{Table_SMTA$^{2}$_pgdli_city}
\end{table*}

\begin{table*}[ht]
\caption{Results of non-stealthy attacks and SMTA$^{2}$ on \textbf{NYUv2} dataset under \textbf{PGD $L_2$} attack. The targeted-task results are underlined. \textcolor{blue}{Blue} and \textcolor{red}{red} indicate affected and preserved values, respectively. (↑)/(↓) denote higher/lower is better.}
\centering
\resizebox{\linewidth}{!}{
\begin{tabular}{c|ccc|ccc|ccc}
\toprule
\begin{tabular}
{@{}c@{}} Method \end{tabular} & \multicolumn{3}{c|}{Non-stealthy} & \multicolumn{3}{c|}{SMTA$^{2}$ Manual} & \multicolumn{3}{c}{SMTA$^{2}$ Auto} \\ \midrule
\begin{tabular}
{@{}c@{}} Tasks \end{tabular} & \begin{tabular}[c]{@{}c@{}}Segment\\ \begin{small}{[}mIoU(↑){]}\end{small}\end{tabular} & \begin{tabular}[c]{@{}c@{}}Depth\\ \begin{small}{[}aErr(↓){]}\end{small}\end{tabular} & \begin{tabular}[c]{@{}c@{}}Normal\\ \begin{small}{[}mDist(↓){]}\end{small}\end{tabular} & \begin{tabular}[c]{@{}c@{}}Segment\\ \begin{small}{[}mIoU(↑){]}\end{small}\end{tabular} & \begin{tabular}[c]{@{}c@{}}Depth\\ \begin{small}{[}aErr(↓){]}\end{small}\end{tabular} & \begin{tabular}[c]{@{}c@{}}Normal\\ \begin{small}{[}mDist(↓){]}\end{small}\end{tabular} & \begin{tabular}[c]{@{}c@{}}Segment\\ \begin{small}{[}mIoU(↑){]}\end{small}\end{tabular} & \begin{tabular}[c]{@{}c@{}}Depth\\ \begin{small}{[}aErr(↓){]}\end{small}\end{tabular} & \begin{tabular}[c]{@{}c@{}}Normal\\ \begin{small}{[}mDist(↓){]}\end{small}\end{tabular}
\\ \midrule
$\epsilon$=0 & 46.56  & 40.57                           & 23.41        & 46.56  & 40.57                           & 23.41  & 46.56  & 40.57                           & 23.41  \\ \midrule
 & \textcolor{blue}{\underline{14.70}}   & \textcolor{blue}{70.56}                        & \textcolor{blue}{33.29} & \textcolor{blue}{\underline{19.30}}   & \textcolor{red}{40.18}    & \textcolor{red}{23.25}  & \textcolor{blue}{\underline{19.44}}                        & \textcolor{red}{37.65}                &  \textcolor{red}{23.35}  \\
$\epsilon$=5   &   \textcolor{blue}{21.04}  & \textcolor{blue}{\underline{197.61}}   & 
 \textcolor{blue}{36.48}  &   \textcolor{red}{46.71}  & \textcolor{blue}{\underline{174.81}}   & 
 \textcolor{red}{23.26} & \textcolor{red}{46.74}   & \textcolor{blue}{\underline{174.57}}   &  \textcolor{red}{22.88}  \\   &  \textcolor{blue}{24.29}   &  \textcolor{blue}{81.77}    &   \textcolor{blue}{\underline{63.31}} &  \textcolor{red}{50.21}   &  \textcolor{red}{40.01}    &   \textcolor{blue}{\underline{71.54}}  &  \textcolor{red}{47.15}   &  \textcolor{red}{39.37}   &  \textcolor{blue}{\underline{54.28}}  \\
\midrule
&  \textcolor{blue}{\underline{7.70}}  &  \textcolor{blue}{87.83}  &  \textcolor{blue}{39.33}  &  \textcolor{blue}{\underline{13.03}}  &  \textcolor{red}{39.78}  &  \textcolor{red}{23.19} & \textcolor{blue}{\underline{12.84}}
& \textcolor{red}{38.05}  &  \textcolor{red}{23.22}    \\  $\epsilon$=10    & \textcolor{blue}{11.20} 
 &  \textcolor{blue}{\underline{290.69}} 
 &  \textcolor{blue}{44.18}   & \textcolor{red}{46.79} 
 &  \textcolor{blue}{\underline{251.49}}  
 &  \textcolor{red}{23.30}  & \textcolor{red}{46.68} 
  &  \textcolor{blue}{\underline{251.15}} 
  &   \textcolor{red}{22.56}       \\            & \textcolor{blue}{14.18}   &   \textcolor{blue}{107.80}   &   \textcolor{blue}{\underline{81.43}} & \textcolor{red}{47.79}   &   \textcolor{red}{39.22}   &   \textcolor{blue}{\underline{71.54}}  &  \textcolor{red}{49.71}  &  \textcolor{red}{38.16} & \textcolor{blue}{\underline{68.21}}     \\ 
\bottomrule
\end{tabular}}
\label{Table_SMTA$^{2}$_pgdl2_nyuv2}
\end{table*}
\subsection{Main Results and Analysis}

\begin{table*}[!t]
\caption{Results of non-stealthy attacks and SMTA$^{2}$ on \textbf{Cityscapes} dataset under \textbf{PGD $L_2$} attack. The targeted-task results are underlined. \textcolor{blue}{Blue} and \textcolor{red}{red} indicate affected and preserved values, respectively. (↑)/(↓) denote higher/lower is better.}
\centering
\resizebox{\linewidth}{!}{
\begin{tabular}{c|ccc|ccc|ccc}
\toprule
\begin{tabular}
{@{}c@{}} Method \end{tabular} & \multicolumn{3}{c|}{Non-stealthy} & \multicolumn{3}{c|}{SMTA$^{2}$ Manual} & \multicolumn{3}{c}{SMTA$^{2}$ Auto} \\ \midrule
\begin{tabular}
{@{}c@{}} Tasks\end{tabular} & \begin{tabular}[c]{@{}c@{}}Segment\\ \begin{small}{[}mIoU(↑){]}\end{small}\end{tabular} & \begin{tabular}[c]{@{}c@{}}Part Seg\\ \begin{small}{[}mIoU(↑){]}\end{small}\end{tabular} & \begin{tabular}[c]{@{}c@{}}Disparity\\ \begin{small}{[}aErr(↓){]}\end{small}\end{tabular} &
\begin{tabular}[c]{@{}c@{}}Segment\\ \begin{small}{[}mIoU(↑){]}\end{small}\end{tabular} & \begin{tabular}[c]{@{}c@{}}Part Seg\\ \begin{small}{[}mIoU(↑){]}\end{small}\end{tabular} & \begin{tabular}[c]{@{}c@{}}Disparity\\ \begin{small}{[}aErr(↓){]}\end{small}\end{tabular} & \begin{tabular}[c]{@{}c@{}}Segment\\ \begin{small}{[}mIoU(↑){]}\end{small}\end{tabular} & \begin{tabular}[c]{@{}c@{}}Part Seg\\ \begin{small}{[}mIoU(↑){]}\end{small}\end{tabular} & \begin{tabular}[c]{@{}c@{}}Disparity\\ \begin{small}{[}aErr(↓){]}\end{small}\end{tabular} \\ \midrule
$\epsilon$=0 & 54.20  & 51.82             & 81.51        & 54.20  & 51.82             & 81.51  & 54.20  & 51.82             & 81.51 \\ \midrule
 & \textcolor{blue}{\underline{18.84}}   & \textcolor{blue}{33.17}                        & \textcolor{blue}{126.93} & \textcolor{blue}{\underline{21.48}}   & \textcolor{red}{53.70}                        & \textcolor{red}{80.36}  & \textcolor{blue}{\underline{21.20}}                        & \textcolor{red}{52.36}                &  \textcolor{red}{81.44}  \\
$\epsilon$=5   &   \textcolor{blue}{31.08}  & \textcolor{blue}{\underline{16.60}}   & 
 \textcolor{blue}{144.51}  &   \textcolor{red}{55.72}  & \textcolor{blue}{\underline{19.56}}   & 
 \textcolor{red}{80.21}   &  \textcolor{red}{56.49}   & \textcolor{blue}{\underline{19.71}}  &  \textcolor{red}{81.32}  \\   &  \textcolor{blue}{22.03}   &  \textcolor{blue}{17.73}    &   \textcolor{blue}{\underline{921.67}} &  \textcolor{blue}{56.39}   &  \textcolor{blue}{53.45}    &   \textcolor{blue}{\underline{652.57}}  &  \textcolor{blue}{57.69}   &  \textcolor{blue}{53.47}   &  \textcolor{blue}{\underline{607.56}}  \\
\midrule
&  \textcolor{blue}{\underline{10.52}}  &  \textcolor{blue}{17.98} &  \textcolor{blue}{167.64} &  \textcolor{blue}{\underline{14.85}}  &  \textcolor{red}{55.68}  &  \textcolor{red}{81.43}  & \textcolor{blue}{\underline{14.25}} 
& \textcolor{red}{52.99}  &  \textcolor{red}{80.35}  \\    $\epsilon$=10    & \textcolor{blue}{22.95} 
 &  \textcolor{blue}{\underline{9.86}}  
 &  \textcolor{blue}{274.18}    & \textcolor{red}{55.58} 
 &  \textcolor{blue}{\underline{13.70}}  
 &  \textcolor{red}{81.18}    & \textcolor{red}{56.11} 
  &  \textcolor{blue}{\underline{13.49}} 
  &   \textcolor{red}{81.39}       \\  
  & \textcolor{blue}{12.93}  & \textcolor{blue}{8.33}  &  \textcolor{blue}{\underline{1611.60}} & \textcolor{red}{55.24}   &   \textcolor{red}{52.00}   &   \textcolor{blue}{\underline{1057.77}}  &  \textcolor{red}{57.53}  &  \textcolor{red}{51.99} & \textcolor{blue}{\underline{1026.31}}     \\ 
\bottomrule
\end{tabular}}
\label{Table_SMTA$^{2}$_pgdl2_city}
\end{table*}

\subsubsection{Attack Results of the Undefended Model}

\begin{table}[htbp]
\centering
\caption{Experimental results of SMTA$^{2}$ Auto method with APGD attacks with $L_1$ \& $L_2$ norms on Cityscapes. The targeted results are underlined. \textcolor{blue}{Blue} and \textcolor{red}{red} indicate affected and preserved values, respectively. (↑)/(↓) denote higher/lower is better.}
\centering
\resizebox{0.7\linewidth}{!}{\begin{tabular}{c|ccc|c|ccc}
\toprule
APGD $L_1$ & \textbf{Seg}($\uparrow$) & \textbf{Part}($\uparrow$) & \textbf{Disp}($\downarrow$) & APGD $L_2$ & \textbf{Seg}($\uparrow$) & \textbf{Part}($\uparrow$) & \textbf{Disp}($\downarrow$) \\
\midrule
$\epsilon$=0 & 54.20 & 51.82 & 81.51 & $\epsilon$=0 & 54.20 & 51.82 & 81.51 \\
\midrule
\multirow{3}{*}{$\epsilon$=10}
& \textcolor{blue}{\underline{18.84}} & \textcolor{red}{51.88} & \textcolor{red}{78.60}
&  & \textcolor{blue}{\underline{23.21}} & \textcolor{red}{51.98} & \textcolor{red}{80.76} \\
& \textcolor{red}{55.69} & \textcolor{blue}{\underline{19.90}} & \textcolor{red}{80.16}
& $\epsilon$=5  & \textcolor{red}{54.44} & \textcolor{blue}{\underline{24.08}} & \textcolor{red}{81.26} \\
& \textcolor{red}{55.85} & \textcolor{red}{52.48} & \textcolor{blue}{\underline{461.82}}
& & \textcolor{red}{54.87} & \textcolor{red}{52.71} & \textcolor{blue}{\underline{387.33}}  \\
\midrule

\multirow{3}{*}{$\epsilon$=20}
& \textcolor{blue}{\underline{13.55}} & \textcolor{red}{52.84} & \textcolor{red}{80.64}
&  & \textcolor{blue}{\underline{15.03}} & \textcolor{red}{52.53} & \textcolor{red}{81.10} \\
& \textcolor{red}{55.28} & \textcolor{blue}{\underline{13.58}} & \textcolor{red}{80.23}
& $\epsilon$=10 & \textcolor{red}{56.20} & \textcolor{blue}{\underline{14.12}} & \textcolor{red}{80.36} \\
& \textcolor{red}{56.74} & \textcolor{red}{53.04} & \textcolor{blue}{\underline{827.69}}
&  & \textcolor{red}{55.99} & \textcolor{red}{52.08} & \textcolor{blue}{\underline{690.18}} \\
\bottomrule
\end{tabular}}
\label{APGD}
\end{table}

Results on NYUv2 under PGD $L_\infty$ attacks (Table~\ref{Table_SMTA$^{2}$_pgdli_nyuv2}) show that non-stealthy attacks degrade all tasks, whereas both manual and automated SMTA$^{2}$ selectively degrade the targeted task while preserving non-targeted tasks at baseline or better performance. Even at small perturbation levels ($\epsilon=2/255$ or $4/255$), SMTA$^{2}$ achieves attack strength comparable to non-stealthy methods. As $\epsilon$ increases, the performance gap remains limited, indicating stable and robust selective attack behavior.

The results using Cityscapes dataset (Table~\ref{Table_SMTA$^{2}$_pgdli_city}) exhibit consistent trends. SMTA$^{2}$ maintains non-targeted task performance while achieving strong targeted degradation, often matching or surpassing non-stealthy baselines. The automated weight search performs comparably to manual tuning with significantly higher efficiency, demonstrating scalability and practicality.

\begin{figure*}[!ht]
\centering
\includegraphics[width=\linewidth]{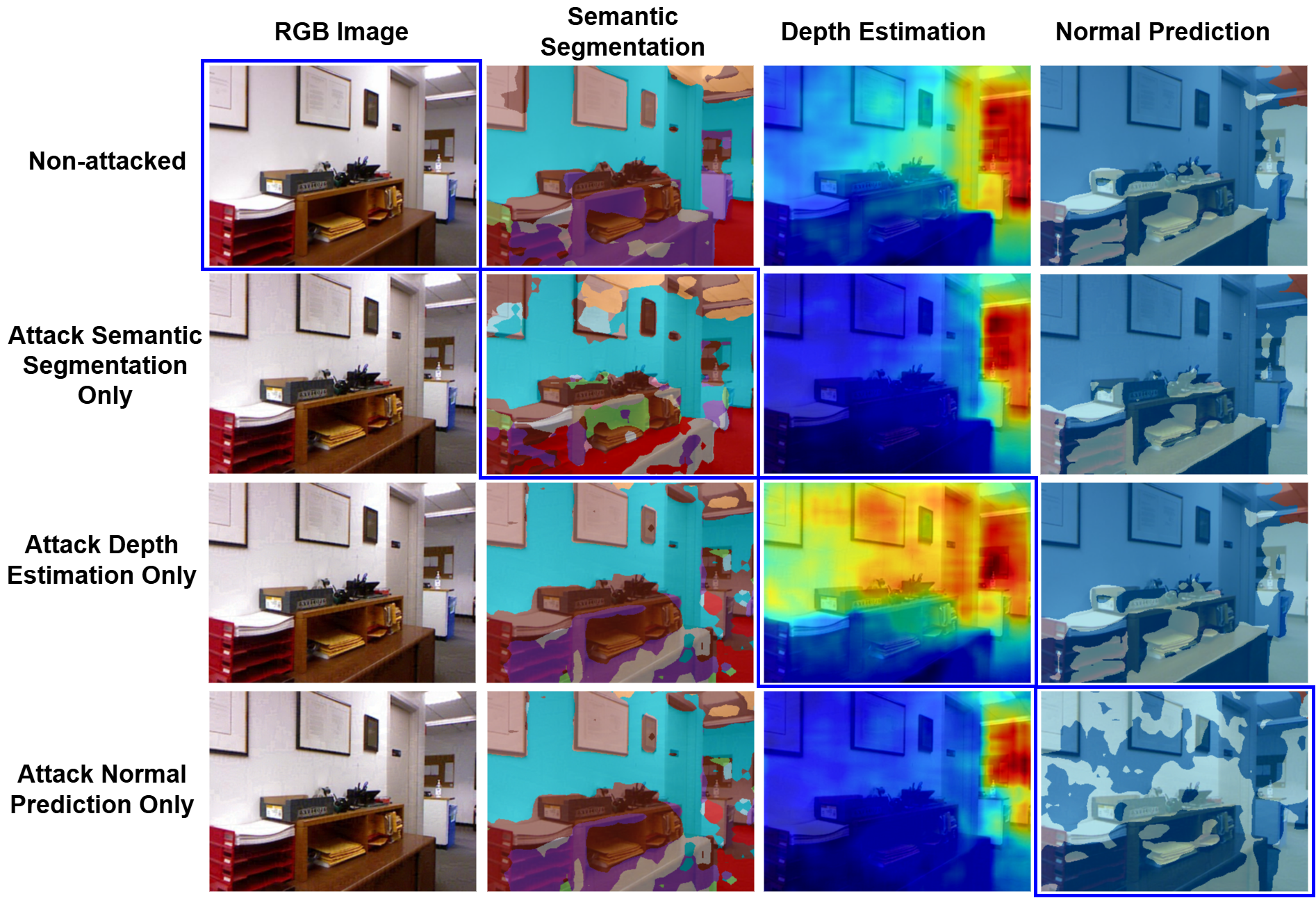} 
\caption{Visualization results of non-attacked outputs and SMTA$^{2}$ attacks on NYUv2 dataset ($\epsilon$=4/255). Each row corresponds to a targeted task, while columns show the targeted and non-targeted task outputs. SMTA$^{2}$ effectively degrades the targeted task while preserving non-targeted tasks.}
\label{Atk_nyuv2_2}
\end{figure*}

\begin{figure*}[!ht]
\centering
\includegraphics[width=\linewidth]{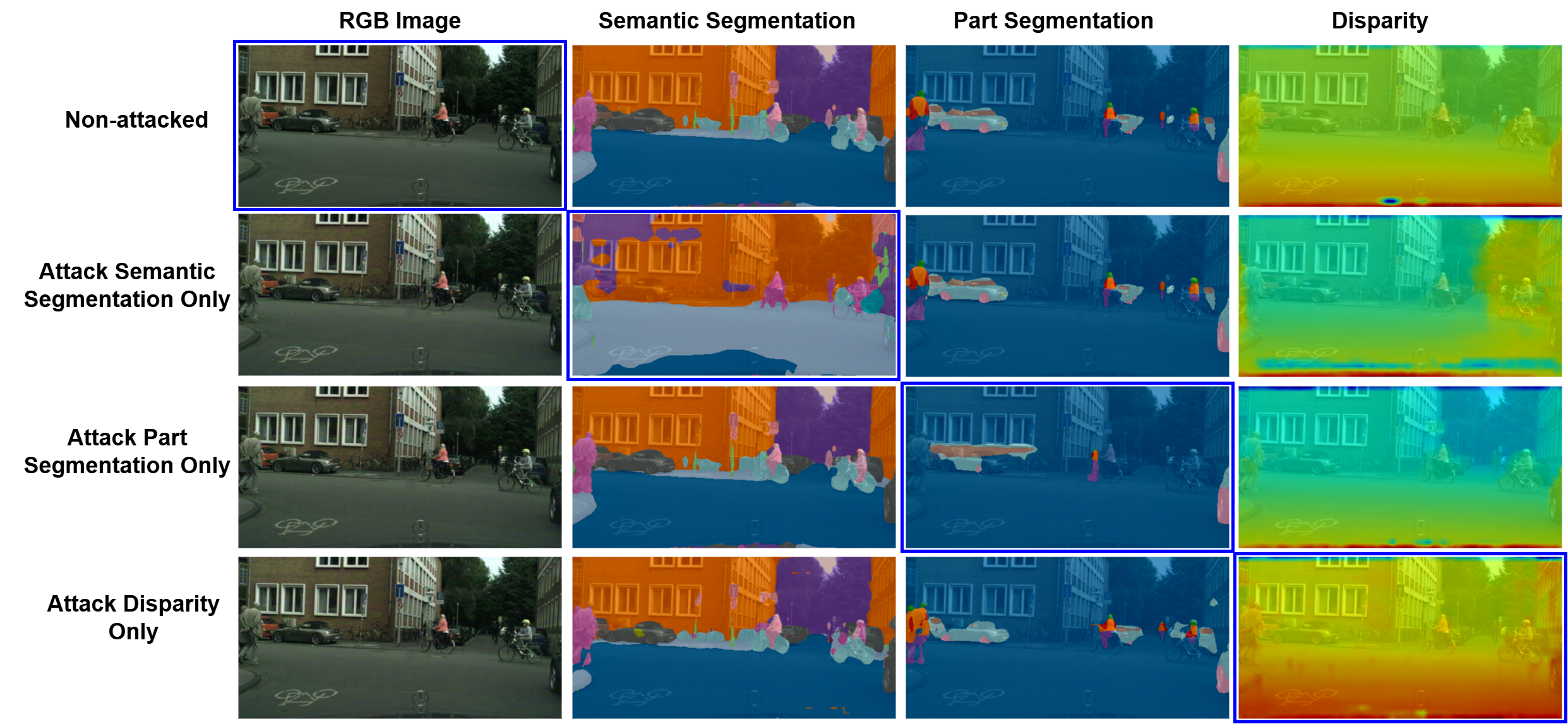} 
\caption{Visualization results of non-attacked outputs and SMTA$^{2}$ attacks on Cityscapes dataset ($\epsilon$=4/255). Each row corresponds to a targeted task, while columns show the targeted and non-targeted task outputs..}
\label{Atk_city_2}
\end{figure*}

Figures~\ref{Atk_nyuv2_2} and~\ref{Atk_city_2} visualize the results of stealthy adversarial attacks on the NYUv2 and Cityscapes datasets with $\epsilon$=4/255, respectively. The perturbations introduce only subtle pixel-level changes, yet they significantly degrade the performance of the targeted task while effectively preserving the outputs of non-targeted tasks. These examples demonstrate that SMTA$^{2}$ successfully achieves selective task degradation without introducing noticeable visual artifacts, particularly for safety-critical objects such as vehicles and pedestrians.

The SMTA$^{2}$ framework is further evaluated under PGD-$L_2$ and stronger APGD attacks with both $L_1$ and $L_2$ norms. As shown in Tables~\ref{Table_SMTA$^{2}$_pgdl2_nyuv2}, \ref{Table_SMTA$^{2}$_pgdl2_city}, and~\ref{APGD}, SMTA$^{2}$ consistently achieves effective task-selective attacks across different perturbation budgets and attack algorithms. The targeted task is significantly degraded, while non-targeted tasks remain hold clean performance. Moreover, the automated searching generally matches or outperforms manual tuning. These results demonstrate that SMTA$^{2}$ generalizes well across different attack paradigms and norm constraints, maintaining strong attack effectiveness and stealthy task preservation under multiple attack algorithms.

\subsubsection{Visualized Results of Attack Effectiveness}

\begin{figure}[!ht]
\centering
\includegraphics[width=\linewidth]{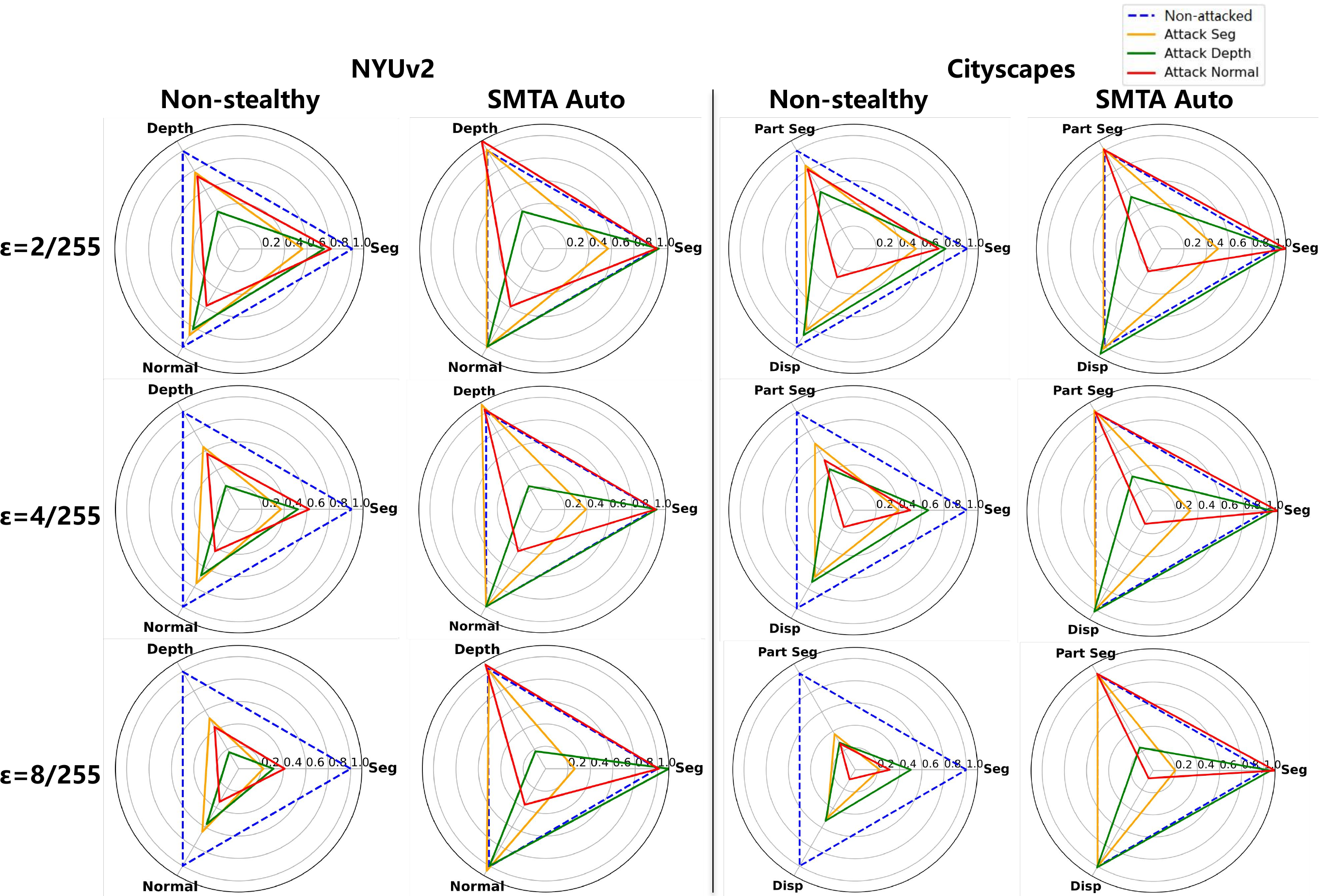} 
\caption{Radar images of undefended PGD $L_\infty$ attack effectiveness of non-stealthy and SMTA$^{2}$ Automated methods of NYUv2 and Cityscapes datasets.}
\label{radar_pgdlinf_undefended}
\end{figure}

Figure~\ref{radar_pgdlinf_undefended} visualizes the attack effectiveness of non-stealthy and automated SMTA$^{2}$ methods under PGD $L_\infty$ attacks using radar charts, which depict the ratio between post-attack and pre-attack performance. For metrics where higher values indicate better performance (e.g., mIoU), the ratio is computed as post-attack divided by pre-attack performance, while for metrics where lower values are preferable (e.g., aErr), the inverse ratio is used. Smaller ratios (closer to the center) indicate stronger attack effectiveness, whereas larger ratios (closer to the boundary) correspond to better retained performance. The dashed blue line denotes the clean baseline with all ratios equal to 1.

As shown, non-stealthy attacks significantly degrade all tasks, even when only a single task is targeted, indicating substantial collateral damage. In contrast, automated SMTA$^{2}$ selectively degrades the targeted task while strictly preserving, or even improving, the performance of non-targeted tasks, demonstrating its strong task-selective attack capability and stealthiness.

\subsubsection{Attack Results of Adversarially Trained Models}

\begin{table*}[!ht]
\caption{Results of non-stealthy attacks and SMTA$^{2}$ under PGD adversarial training on \textbf{NYUv2} dataset using \textbf{PGD $L_\infty$} attack. Non-attacked performance is denoted by $\epsilon$=0. The targeted-task results are underlined. \textcolor{blue}{Blue} and \textcolor{red}{red} indicate affected and preserved values, respectively. (↑)/(↓) denote higher/lower is better.}
\centering
\resizebox{\linewidth}{!}{
\begin{tabular}{c|ccc|ccc|ccc}
\toprule
\begin{tabular}
{@{}c@{}} Method \end{tabular} & \multicolumn{3}{c|}{Non-stealthy} & \multicolumn{3}{c|}{SMTA$^{2}$ Manual} & \multicolumn{3}{c}{SMTA$^{2}$ Auto} \\ \midrule
\begin{tabular}
{@{}c@{}} Tasks \end{tabular} & \begin{tabular}[c]{@{}c@{}}Segment\\ \begin{small}{[}mIoU(↑){]}\end{small}\end{tabular} & \begin{tabular}[c]{@{}c@{}}Depth\\ \begin{small}{[}aErr(↓){]}\end{small}\end{tabular} & \begin{tabular}[c]{@{}c@{}}Normal\\ \begin{small}{[}mDist(↓){]}\end{small}\end{tabular} & \begin{tabular}[c]{@{}c@{}}Segment\\ \begin{small}{[}mIoU(↑){]}\end{small}\end{tabular} & \begin{tabular}[c]{@{}c@{}}Depth\\ \begin{small}{[}aErr(↓){]}\end{small}\end{tabular} & \begin{tabular}[c]{@{}c@{}}Normal\\ \begin{small}{[}mDist(↓){]}\end{small}\end{tabular} & \begin{tabular}[c]{@{}c@{}}Segment\\ \begin{small}{[}mIoU(↑){]}\end{small}\end{tabular} & \begin{tabular}[c]{@{}c@{}}Depth\\ \begin{small}{[}aErr(↓){]}\end{small}\end{tabular} & \begin{tabular}[c]{@{}c@{}}Normal\\ \begin{small}{[}mDist(↓){]}\end{small}\end{tabular} \\ \midrule
$\epsilon$=0 & 46.56  & 40.57                           & 23.41        & 46.56  & 40.57                           & 23.41   & 46.56  & 40.57                           & 23.41   \\ \midrule
  & \textcolor{blue}{\underline{33.21}}                        & \textcolor{blue}{52.27}                &  \textcolor{blue}{28.00} & \textcolor{blue}{\underline{35.67}}   & \textcolor{red}{36.82}                        & \textcolor{red}{23.00} & \textcolor{blue}{\underline{33.66}}   & \textcolor{red}{39.94}                        & \textcolor{red}{23.36}         \\
$\epsilon$=2/255    &  \textcolor{blue}{31.00}   & \textcolor{blue}{\underline{61.97}}  &  \textcolor{blue}{30.03}  & \textcolor{red}{47.40}   & \textcolor{blue}{\underline{54.65}}   & 
 \textcolor{red}{23.01}  &   \textcolor{red}{47.39}   & \textcolor{blue}{\underline{56.05}}  & 
\textcolor{red}{22.41}  \\  &  \textcolor{blue}{31.04}   &  \textcolor{blue}{58.55}   & \textcolor{blue}{\underline{31.12}}   &  \textcolor{red}{49.54}  &  \textcolor{red}{40.17}   &   \textcolor{blue}{\underline{32.82}}  &  \textcolor{red}{47.79}   &  \textcolor{red}{38.76}   &   \textcolor{blue}{\underline{32.87}}  \\ 
 \midrule
& \textcolor{blue}{\underline{27.37}} 
 & \textcolor{blue}{59.42}  & \textcolor{blue}{30.28}     &  \textcolor{blue}{\underline{30.91}}  &  \textcolor{red}{39.26}  &  \textcolor{red}{23.37}  &  \textcolor{blue}{\underline{30.99}}  &  \textcolor{red}{38.98}  &  \textcolor{red}{23.31}    \\ $\epsilon$=4/255   &  \textcolor{blue}{34.58}
  &  \textcolor{blue}{\underline{63.36}} 
  &  \textcolor{blue}{28.46}      & \textcolor{red}{47.87} 
 &  \textcolor{blue}{\underline{59.98}}  
&  \textcolor{red}{23.37}     & \textcolor{red}{48.13} 
&  \textcolor{blue}{\underline{61.50}}  
&  \textcolor{red}{23.20}      \\     &  \textcolor{blue}{30.41}  & \textcolor{blue}{59.41}   & \textcolor{blue}{\underline{32.59}}   & \textcolor{red}{46.84}   &   \textcolor{red}{40.64}   &   \textcolor{blue}{\underline{31.06}}  & \textcolor{red}{46.76}   &   \textcolor{red}{39.17}   &   \textcolor{blue}{\underline{30.73}}    \\ \midrule
&  \textcolor{blue}{\underline{23.33}}  &  \textcolor{blue}{61.37}  &  \textcolor{blue}{30.97}  &  \textcolor{blue}{\underline{27.45}}  &   \textcolor{red}{39.87}   &  \textcolor{red}{22.86}  &  \textcolor{blue}{\underline{26.83}} &   \textcolor{red}{38.67}   &  \textcolor{red}{23.17} \\  $\epsilon$=8/255  & \textcolor{blue}{28.68} 
&  \textcolor{blue}{\underline{76.32}}  &  \textcolor{blue}{31.15}     &    \textcolor{red}{48.38} 
&  \textcolor{blue}{\underline{69.60}}  & \textcolor{red}{22.37}   &    \textcolor{red}{46.78} 
&  \textcolor{blue}{\underline{73.72}}  & \textcolor{red}{23.01} \\   & \textcolor{blue}{28.90}  &   \textcolor{blue}{61.44}   &   \textcolor{blue}{\underline{35.85}}  &   \textcolor{red}{48.00}  &  \textcolor{red}{39.95}  & \textcolor{blue}{\underline{35.27}} &   \textcolor{red}{47.38}  &  \textcolor{red}{39.76}  & \textcolor{blue}{\underline{34.42}}     \\
\bottomrule
\end{tabular}}
\label{Table_SMTA$^{2}$_at_pgdli_nyuv2}
\end{table*}

\begin{table*}[!ht]
\caption{Results of non-stealthy attacks and SMTA$^{2}$ under PGD adversarial training on \textbf{Cityscapes} dataset using \textbf{PGD $L_\infty$} attack. The targeted-task results are underlined. \textcolor{blue}{Blue} and \textcolor{red}{red} indicate affected and preserved values, respectively. (↑)/(↓) denote higher/lower is better.}
\centering
\resizebox{\linewidth}{!}{
\begin{tabular}{c|ccc|ccc|ccc}
\toprule
\begin{tabular}
{@{}c@{}} Method \end{tabular} & \multicolumn{3}{c|}{Non-stealthy} & \multicolumn{3}{c|}{SMTA$^{2}$ Manual} & \multicolumn{3}{c}{SMTA$^{2}$ Auto}\\ \midrule
\begin{tabular}
{@{}c@{}} Tasks \end{tabular} & \begin{tabular}[c]{@{}c@{}}Segment\\ \begin{small}{[}mIoU(↑){]}\end{small}\end{tabular} & \begin{tabular}[c]{@{}c@{}}Part Seg\\ \begin{small}{[}mIoU(↑){]}\end{small}\end{tabular} & \begin{tabular}[c]{@{}c@{}}Disparity\\ \begin{small}{[}aErr(↓){]}\end{small}\end{tabular} &
\begin{tabular}[c]{@{}c@{}}Segment\\ \begin{small}{[}mIoU(↑){]}\end{small}\end{tabular} & \begin{tabular}[c]{@{}c@{}}Part Seg\\ \begin{small}{[}mIoU(↑){]}\end{small}\end{tabular} & \begin{tabular}[c]{@{}c@{}}Disparity\\ \begin{small}{[}aErr(↓){]}\end{small}\end{tabular} & \begin{tabular}[c]{@{}c@{}}Segment\\ \begin{small}{[}mIoU(↑){]}\end{small}\end{tabular} & \begin{tabular}[c]{@{}c@{}}Part Seg\\ \begin{small}{[}mIoU(↑){]}\end{small}\end{tabular} & \begin{tabular}[c]{@{}c@{}}Disparity\\ \begin{small}{[}aErr(↓){]}\end{small}\end{tabular} \\ \midrule
$\epsilon$=0  & 54.20   &   51.82                         & 81.51  & 54.20   &   51.82                         & 81.51    & 54.20   &   51.82                         & 81.51 \\ \midrule
 & \textcolor{blue}{\underline{33.21}}            & \textcolor{blue}{38.61}                &  \textcolor{blue}{101.67}  & \textcolor{blue}{\underline{42.26}}   & \textcolor{red}{52.01}                        & \textcolor{red}{80.69} & \textcolor{blue}{\underline{42.09}}  & \textcolor{red}{52.00}                        & \textcolor{red}{81.23}        \\
$\epsilon$=2/255   &  \textcolor{blue}{40.59}  & \textcolor{blue}{\underline{41.09}}  &  \textcolor{blue}{94.69}   & \textcolor{red}{55.12}& \textcolor{blue}{\underline{43.61}}   &  \textcolor{red}{80.68} &   \textcolor{red}{55.20}& \textcolor{blue}{\underline{43.18}}   & 
\textcolor{red}{80.99} \\  & \textcolor{blue}{33.69}   &  \textcolor{blue}{37.34}   &  \textcolor{blue}{\underline{103.34}}  & 
\textcolor{red}{54.78}   &  \textcolor{red}{52.19}   &   \textcolor{blue}{\underline{98.89}}  &  \textcolor{red}{54.83}   &  \textcolor{red}{52.52}   &   \textcolor{blue}{\underline{99.63}}    \\ 
 \midrule
& \textcolor{blue}{\underline{30.98}} & \textcolor{blue}{37.93} &  \textcolor{blue}{102.75}     &  \textcolor{blue}{\underline{30.29}}  &  \textcolor{red}{52.45}  &  \textcolor{red}{81.41}  &  \textcolor{blue}{\underline{31.22}}  &  \textcolor{red}{53.10}  &  \textcolor{red}{81.30}   \\ $\epsilon$=4/255 &  \textcolor{blue}{34.17}
  &  \textcolor{blue}{\underline{34.78}} 
  &  \textcolor{blue}{101.93}        & \textcolor{red}{54.78} 
 &  \textcolor{blue}{\underline{30.82}}  
&  \textcolor{red}{81.22}     & \textcolor{red}{54.96} 
 &  \textcolor{blue}{\underline{31.16}}  
 &  \textcolor{red}{80.66}    \\           & \textcolor{blue}{34.41}   & \textcolor{blue}{38.24}   & \textcolor{blue}{\underline{112.85}}     & \textcolor{red}{54.68}   &   \textcolor{red}{51.94}   &   \textcolor{blue}{\underline{115.52}}  & \textcolor{red}{55.34}   &   \textcolor{red}{52.33}   &   \textcolor{blue}{\underline{114.18}}  \\ \midrule
&  \textcolor{blue}{\underline{27.12}}  &  \textcolor{blue}{36.25}  &  \textcolor{blue}{105.51}  & 
 \textcolor{blue}{\underline{32.56}}  &   \textcolor{red}{53.10}   &  \textcolor{red}{81.21}  &  \textcolor{blue}{\underline{28.69}}  &   \textcolor{red}{52.79}   &  \textcolor{red}{81.08} \\  $\epsilon$=8/255  & \textcolor{blue}{32.32} 
&  \textcolor{blue}{\underline{30.38}}   &  \textcolor{blue}{103.75}    &    \textcolor{red}{54.77}
&  \textcolor{blue}{\underline{33.95}}  & \textcolor{red}{81.04}   &    \textcolor{red}{56.05} 
&  \textcolor{blue}{\underline{30.62}}  & \textcolor{red}{80.21}  \\   & \textcolor{blue}{32.80}  &   \textcolor{blue}{36.69}   &   \textcolor{blue}{\underline{127.97}}     &   \textcolor{red}{54.67}  &  \textcolor{red}{51.97}  & \textcolor{blue}{\underline{112.91}}  &   \textcolor{red}{56.18}  &  \textcolor{red}{52.56}  & \textcolor{blue}{\underline{122.42}} \\
\bottomrule
\end{tabular}}
\label{Table_SMTA$^{2}$_at_pgdli_city}
\end{table*}

\begin{table*}[!ht]
\caption{Results of non-stealthy attacks and SMTA$^{2}$ under PGD adversarial training on \textbf{NYUv2} dataset using \textbf{PGD $L_2$} attack. The targeted-task results are underlined. \textcolor{blue}{Blue} and \textcolor{red}{red} indicate affected and preserved values, respectively. (↑)/(↓) denote higher/lower is better.}
\centering
\resizebox{\linewidth}{!}{
\begin{tabular}{c|ccc|ccc|ccc}
\toprule
\begin{tabular}
{@{}c@{}} Method \end{tabular} & \multicolumn{3}{c|}{Non-stealthy} & \multicolumn{3}{c|}{SMTA$^{2}$ Manual} & \multicolumn{3}{c}{SMTA$^{2}$ Auto} \\ \midrule
\begin{tabular}
{@{}c@{}} Tasks \end{tabular} & \begin{tabular}[c]{@{}c@{}}Segment\\ \begin{small}{[}mIoU(↑){]}\end{small}\end{tabular} & \begin{tabular}[c]{@{}c@{}}Depth\\ \begin{small}{[}aErr(↓){]}\end{small}\end{tabular} & \begin{tabular}[c]{@{}c@{}}Normal\\ \begin{small}{[}mDist(↓){]}\end{small}\end{tabular} & \begin{tabular}[c]{@{}c@{}}Segment\\ \begin{small}{[}mIoU(↑){]}\end{small}\end{tabular} & \begin{tabular}[c]{@{}c@{}}Depth\\ \begin{small}{[}aErr(↓){]}\end{small}\end{tabular} & \begin{tabular}[c]{@{}c@{}}Normal\\ \begin{small}{[}mDist(↓){]}\end{small}\end{tabular} & \begin{tabular}[c]{@{}c@{}}Segment\\ \begin{small}{[}mIoU(↑){]}\end{small}\end{tabular} & \begin{tabular}[c]{@{}c@{}}Depth\\ \begin{small}{[}aErr(↓){]}\end{small}\end{tabular} & \begin{tabular}[c]{@{}c@{}}Normal\\ \begin{small}{[}mDist(↓){]}\end{small}\end{tabular}
\\ \midrule
$\epsilon$=0 & 46.56  & 40.57   &              23.41        & 46.56  & 40.57  &               23.41  & 46.56  & 40.57  &  23.41  \\ \midrule
 & \textcolor{blue}{\underline{25.65}}   & \textcolor{blue}{60.49}                        & \textcolor{blue}{30.72} & \textcolor{blue}{\underline{34.25}}   & \textcolor{red}{40.25}                        & \textcolor{red}{23.02}  & \textcolor{blue}{\underline{34.20}}                        & \textcolor{red}{40.33}                &  \textcolor{red}{23.26}  \\
$\epsilon$=5   &  \textcolor{blue}{29.26}   &\textcolor{blue}{\underline{71.26}}   & \textcolor{blue}{30.96}  &   \textcolor{red}{46.83}  &\textcolor{blue}{\underline{61.80}}   &  \textcolor{red}{23.34} & \textcolor{red}{46.66}   & \textcolor{blue}{\underline{61.97}}   &  \textcolor{red}{22.94}  \\   & \textcolor{blue}{29.32}    & \textcolor{blue}{60.86}     &   \textcolor{blue}{\underline{34.46}} &  \textcolor{red}{46.93}   &  \textcolor{red}{39.69}    &   \textcolor{blue}{\underline{31.52}}  &  \textcolor{red}{47.02}   &  \textcolor{red}{40.56}   &  \textcolor{blue}{\underline{35.21}}  \\
\midrule
& \textcolor{blue}{\underline{20.16}}  &  \textcolor{blue}{64.05}  &  \textcolor{blue}{32.24}  &  \textcolor{blue}{\underline{24.88}}  &  \textcolor{red}{39.68}  &  \textcolor{red}{23.20} & \textcolor{blue}{\underline{24.58}} 
& \textcolor{red}{38.93}  &  \textcolor{red}{22.42}    \\  $\epsilon$=10    & \textcolor{blue}{29.97} 
 &  \textcolor{blue}{\underline{88.34}}  
 &  \textcolor{blue}{30.98}   & \textcolor{red}{46.92} 
 &  \textcolor{blue}{\underline{78.81}}  
 &  \textcolor{red}{22.82}  & \textcolor{red}{46.94} 
  & \textcolor{blue}{\underline{83.19}} 
  &   \textcolor{red}{22.37}       \\            &  \textcolor{blue}{26.35}  & \textcolor{blue}{64.80}     & \textcolor{blue}{\underline{40.39}}  & \textcolor{red}{47.00}   &   \textcolor{red}{39.36}   &  \textcolor{blue}{\underline{38.81}}  &  \textcolor{red}{47.25}  &  \textcolor{red}{38.66} & \textcolor{blue}{\underline{38.55}}     \\ 
\bottomrule
\end{tabular}}
\label{Table_SMTA$^{2}$_at_pgdl2_nyuv2}
\end{table*}

\begin{table*}[!ht]
\caption{Results of non-stealthy attacks and SMTA$^{2}$ under PGD adversarial training on \textbf{Cityscapes} dataset using \textbf{PGD $L_2$} attack. The targeted-task results are underlined. \textcolor{blue}{Blue} and \textcolor{red}{red} indicate affected and preserved values, respectively. (↑)/(↓) denote higher/lower is better.}
\centering
\resizebox{\linewidth}{!}{
\begin{tabular}{c|ccc|ccc|ccc}
\toprule
\begin{tabular}
{@{}c@{}} Method \end{tabular} & \multicolumn{3}{c|}{Non-stealthy} & \multicolumn{3}{c|}{SMTA$^{2}$ Manual} & \multicolumn{3}{c}{SMTA$^{2}$ Auto} \\ \midrule
\begin{tabular}
{@{}c@{}} Tasks\end{tabular} & \begin{tabular}[c]{@{}c@{}}Segment\\ \begin{small}{[}mIoU(↑){]}\end{small}\end{tabular} & \begin{tabular}[c]{@{}c@{}}Part Seg\\ \begin{small}{[}mIoU(↑){]}\end{small}\end{tabular} & \begin{tabular}[c]{@{}c@{}}Disparity\\ \begin{small}{[}aErr(↓){]}\end{small}\end{tabular} &
\begin{tabular}[c]{@{}c@{}}Segment\\ \begin{small}{[}mIoU(↑){]}\end{small}\end{tabular} & \begin{tabular}[c]{@{}c@{}}Part Seg\\ \begin{small}{[}mIoU(↑){]}\end{small}\end{tabular} & \begin{tabular}[c]{@{}c@{}}Disparity\\ \begin{small}{[}aErr(↓){]}\end{small}\end{tabular} & \begin{tabular}[c]{@{}c@{}}Segment\\ \begin{small}{[}mIoU(↑){]}\end{small}\end{tabular} & \begin{tabular}[c]{@{}c@{}}Part Seg\\ \begin{small}{[}mIoU(↑){]}\end{small}\end{tabular} & \begin{tabular}[c]{@{}c@{}}Disparity\\ \begin{small}{[}aErr(↓){]}\end{small}\end{tabular} \\ \midrule
$\epsilon$=0 & 54.20  & 51.82 & 81.51        & 54.20  & 51.82 & 81.51  & 54.20  & 51.82 & 81.51 \\ \midrule
 & \textcolor{blue}{\underline{26.03}}   &  \textcolor{blue}{35.76}                       & \textcolor{blue}{108.67} & \textcolor{blue}{\underline{26.28}}   & \textcolor{red}{52.84}                        & \textcolor{red}{81.03}  & {\textcolor{blue}{\underline{26.22}}}                        & \textcolor{red}{52.13}                &  \textcolor{red}{81.44}  \\
$\epsilon$=5   &  \textcolor{blue}{35.67}   &\textcolor{blue}{\underline{33.06}}   & \textcolor{blue}{98.76} &   \textcolor{red}{55.72}  &\textcolor{blue}{\underline{30.26}}   & 
 \textcolor{red}{80.81}   &  \textcolor{red}{56.13}   & \textcolor{blue}{\underline{31.17}}  &  \textcolor{red}{81.11}  \\   &  \textcolor{blue}{36.84}   &   \textcolor{blue}{39.65}   &   \textcolor{blue}{\underline{135.95}} &  \textcolor{red}{56.20}   &  \textcolor{red}{52.11}    &   \textcolor{blue}{\underline{133.21}}  &  \textcolor{red}{56.92}   &  \textcolor{red}{53.41}   &  \textcolor{blue}{\underline{128.56}}  \\
\midrule
&    \textcolor{blue}{\underline{21.06}}  &  \textcolor{blue}{31.83}  &  \textcolor{blue}{119.86}  &  \textcolor{blue}{\underline{29.46}}  &  \textcolor{red}{52.88}  & \textcolor{red}{80.93} & \textcolor{blue}{\underline{28.51}} 
 & \textcolor{red}{52.52}  &  \textcolor{red}{80.78}    \\  $\epsilon$=10    & \textcolor{blue}{26.90}
 &  \textcolor{blue}{\underline{21.37}}  
  &  \textcolor{blue}{116.82}   & \textcolor{red}{54.96} 
  &  \textcolor{blue}{\underline{28.57}}  
  &  \textcolor{red}{81.40}  & \textcolor{red}{55.15}
   & \textcolor{blue}{\underline{26.76}}
   &   \textcolor{red}{80.58}       \\            &  \textcolor{blue}{26.22}  & \textcolor{blue}{27.53}     &  \textcolor{blue}{\underline{229.00}} & \textcolor{red}{56.31}   &   \textcolor{red}{53.46}   &  \textcolor{blue}{\underline{160.88}}  &  \textcolor{red}{56.74}  &  \textcolor{red}{53.40} & \textcolor{blue}{\underline{165.08}}     \\ 
\bottomrule
\end{tabular}}
\label{Table_SMTA$^{2}$_at_pgdl2_city}
\end{table*}

To evaluate robustness under defensive settings, we further conduct experiments on Adversarially Trained (AT) models obtained via PGD-based adversarial training. Tables~\ref{Table_SMTA$^{2}$_at_pgdli_nyuv2} and~\ref{Table_SMTA$^{2}$_at_pgdli_city} present results under PGD $L_\infty$ attacks on NYUv2 and Cityscapes using the same perturbation budgets ($\epsilon$=2/255, 4/255, 8/255). Across all settings, SMTA$^{2}$ consistently preserves non-targeted tasks while effectively degrading the targeted one, demonstrating stable task-selective attack capability. Both manual and automated variants show reliable performance.

Similar trends are observed under PGD $L_2$ attacks (Tables~\ref{Table_SMTA$^{2}$_at_pgdl2_nyuv2} and~\ref{Table_SMTA$^{2}$_at_pgdl2_city}). While non-stealthy attacks degrade all tasks, SMTA$^{2}$ maintains baseline performance for non-targeted tasks and achieves consistent targeted degradation. Despite enhanced robustness from adversarial training, SMTA$^{2}$ remains effective, confirming its reliability under strong defense settings.

\section{Conclusion}

We propose Stealthy Multi-Task Adversarial Attacks (SMTA$^{2}$), a principled framework for selectively degrading a targeted task while strictly preserving non-targeted tasks in multi-task learning. Unlike conventional adversarial attacks that cause indiscriminate performance collapse, SMTA$^{2}$ formulates selective degradation as a constrained multi-objective optimization problem and solves it via adaptive weight optimization. Despite the strong inter-task coupling in shared representations in multi-task learning, SMTA$^{2}$ reliably achieves targeted attack effectiveness without collateral damage, even under adversarial trained scenarios. Experiments demonstrate that SMTA$^{2}$ consistently achieves strong attack effectiveness on targeted tasks while preserving non-targeted tasks across undefended and adversarially trained settings. This work demonstrates that stealthy, task-selective attacks are both feasible and practically solvable, opening a new direction for controlled adversarial machine learning in multi-task systems.

\section*{Acknowledgments}
This research was supported by the National Science Foundation under award CNS-2245765.

%
%
\bibliographystyle{splncs04}
\bibliography{references}
\end{document}



\title{Stealthy Multi-task Adversarial Attacks
\\
--Supplementary Material--} 

\titlerunning{Stealthy Multi-task Adversarial Attacks}

\author{Jiacheng Guo\inst{1,2*}\orcidlink{0009-0006-8508-9949} \and
Tianyun Zhang\inst{1*\dagger}\orcidlink{0000-0002-2475-6414} \and
Lei Li\inst{1}\orcidlink{0009-0003-4317-9071} \and
Haochen Yang\inst{1}\orcidlink{0000-0001-9145-8575} \and
Hongkai Yu\inst{1}\orcidlink{0000-0001-5383-8913} \and
Minghai Qin\inst{1,3\dagger}\orcidlink{0000-0001-5172-5309}}

\authorrunning{J.~Guo et al.}

\institute{Cleveland State University, Cleveland, OH 44115, USA \email{\{j.guo58,l.li15,h.yang15\}@vikes.csuohio.edu}, \email{\{t.zhang85,h.yu19\}@csuohio.edu} \and
University of Wisconsin-Madison, Madison, WI 53706, USA \and
Western Digital Research, San Jose, CA 95119, USA\\ \email{\{minghai.qin\}@wdc.com}\\ 
$^{*}$Equal Contribution $^{\dagger}$Corresponding Authors
}

\maketitle

\section{Extensive Experiment Results}

Besides the displayed Projected Gradient Descent (PGD)~\cite{madry2018towards} attack method, we also evaluates the performance of the proposed SMTA$^{2}$ framework using another attack strategy: Iterative Fast Gradient Sign Method (IFGSM)~\cite{kurakin2018adversarial}. The IFGSM attack is implemented in $L_{\infty}$ norm. The IFGSM~\cite{kurakin2018adversarial} is a special case of PGD that uses the sign of the gradient for each update and typically uses a fixed step size; it's a stronger, multi-step version of the basic Fast Gradient Sign Method (FGSM)~\cite{goodfellow2014explaining}. 

\subsection{Results using IFGSM attack}

Similar conclusions to those reached in the main paper are drawn from Tables \ref{Table_smta_ifgsm_nyuv2} and \ref{Table_smta_ifgsm_city}, where the SMTA$^{2}$ framework's manual and automated methods are applied using the IFGSM attack algorithm. The attack effectiveness using IFGSM is nearly identical to the PGD $L_\infty$ results, further demonstrating that the SMTA$^{2}$ framework is robust and consistent across various attack methods.

\begin{table*}[!ht]
\caption{Experimental results of non-stealthy and SMTA$^{2}$ framework by manual and automated solutions of \textbf{IFGSM} attack algorithm on \textbf{NYUv2} dataset. Values affected by the attack are marked in \textcolor{blue}{blue}, while those not influenced by the attack are highlighted in \textcolor{red}{red}. The performances of the targeted task are underlined. (↑) means higher is better and (↓) means lower is better.}
\centering
\resizebox{\linewidth}{!}{
\begin{tabular}{c|ccc|ccc|ccc}
\toprule
\begin{tabular}
{@{}c@{}} Method \end{tabular} & \multicolumn{3}{c|}{Non-stealthy} & \multicolumn{3}{c|}{SMTA$^{2}$ Manual} & \multicolumn{3}{c}{SMTA$^{2}$ Auto} \\ \midrule
\begin{tabular}
{@{}c@{}} Tasks \end{tabular} & \begin{tabular}[c]{@{}c@{}}Segment\\ \begin{small}{[}mIoU(↑){]}\end{small}\end{tabular} & \begin{tabular}[c]{@{}c@{}}Depth\\ \begin{small}{[}aErr(↓){]}\end{small}\end{tabular} & \begin{tabular}[c]{@{}c@{}}Normal\\ \begin{small}{[}mDist(↓){]}\end{small}\end{tabular} & \begin{tabular}[c]{@{}c@{}}Segment\\ \begin{small}{[}mIoU(↑){]}\end{small}\end{tabular} & \begin{tabular}[c]{@{}c@{}}Depth\\ \begin{small}{[}aErr(↓){]}\end{small}\end{tabular} & \begin{tabular}[c]{@{}c@{}}Normal\\ \begin{small}{[}mDist(↓){]}\end{small}\end{tabular} & \begin{tabular}[c]{@{}c@{}}Segment\\ \begin{small}{[}mIoU(↑){]}\end{small}\end{tabular} & \begin{tabular}[c]{@{}c@{}}Depth\\ \begin{small}{[}aErr(↓){]}\end{small}\end{tabular} & \begin{tabular}[c]{@{}c@{}}Normal\\ \begin{small}{[}mDist(↓){]}\end{small}\end{tabular} \\ \midrule
$\epsilon$=0 & 46.56  & 40.57                           & 23.41        & 46.56  & 40.57                           & 23.41   & 46.56  & 40.57                           & 23.41   \\ \midrule
& {\underline{\textcolor{blue}{26.21}}}                        & \textcolor{blue}{51.77}                &  \textcolor{blue}{26.65} & \underline{\textcolor{blue}{26.42}}   & \textcolor{red}{38.25}                        & \textcolor{red}{23.07}  & \underline{\textcolor{blue}{26.44}}   & \textcolor{red}{36.74}                        & \textcolor{red}{23.33}        \\
$\epsilon$=2/255  &  \textcolor{blue}{35.10}   & {\underline{\textcolor{blue}{106.92}}}  &  \textcolor{blue}{28.45}   & \textcolor{red}{47.43}   &{\underline{\textcolor{blue}{106.49}}}   & 
 \textcolor{red}{23.06}  &   \textcolor{red}{47.62}   &{\underline{\textcolor{blue}{106.33}}}   & 
 \textcolor{red}{23.34} \\  &  \textcolor{blue}{37.76}   &  \textcolor{blue}{54.96}   &  {\underline{\textcolor{blue}{40.36}}}    &  \textcolor{red}{48.27}   &  \textcolor{red}{38.54}   &   {\underline{\textcolor{blue}{41.12}}}  &  \textcolor{red}{46.94}   &  \textcolor{red}{39.50}   &   {\underline{\textcolor{blue}{40.81}}}  \\ 
 \midrule
 & \underline{\textcolor{blue}{17.04}} 
 & \textcolor{blue}{63.80}  &  \textcolor{blue}{30.97}    &  \underline{\textcolor{blue}{18.26}}  &  \textcolor{red}{37.45}  &  \textcolor{red}{23.34}   &  \underline{\textcolor{blue}{18.33}}  &  \textcolor{red}{37.91}  &  \textcolor{red}{23.32}   \\ $\epsilon$=4/255  & \textcolor{blue}{24.15} 
  &  \underline{\textcolor{blue}{167.40}} 
  &   \textcolor{blue}{34.39}     & \textcolor{red}{46.67} 
 &  \underline{\textcolor{blue}{164.99}}  
&  \textcolor{red}{23.05}     & \textcolor{red}{46.72} 
 &  \underline{\textcolor{blue}{169.30}}  
 &  \textcolor{red}{23.22}      \\           &  \textcolor{blue}{28.88}  & \textcolor{blue}{71.44}  & \underline{\textcolor{blue}{53.88}}    & \textcolor{red}{47.97}   &   \textcolor{red}{37.39}   &   \underline{\textcolor{blue}{53.45}}  & \textcolor{red}{51.12}   &  \textcolor{red}{38.85}    &   \underline{\textcolor{blue}{52.55}}   \\
\midrule
&  \underline{\textcolor{blue}{10.11}}  &  \textcolor{blue}{78.36}  &  \textcolor{blue}{36.23}  & 
\underline{\textcolor{blue}{13.01}}  &   \textcolor{red}{38.03}   &  \textcolor{red}{22.67}  &  \underline{\textcolor{blue}{12.42}}  &   \textcolor{red}{39.34}   &  \textcolor{red}{23.26}  
 \\ $\epsilon$=8/255   & \textcolor{blue}{14.58} 
&  \underline{\textcolor{blue}{240.52}}   &  \textcolor{blue}{41.16}    &    \textcolor{red}{46.63} 
&  \underline{\textcolor{blue}{230.47}}  & \textcolor{red}{22.13}   &   \textcolor{red}{46.77}  
&  \underline{\textcolor{blue}{203.15}}  &  \textcolor{red}{23.31} \\   & \textcolor{blue}{18.91}  &   \textcolor{blue}{93.36}   &   \underline{\textcolor{blue}{68.05}}    &   \textcolor{red}{46.66}  &  \textcolor{red}{40.53}  & \underline{\textcolor{blue}{66.72}}   &   \textcolor{red}{49.68}  &  \textcolor{red}{38.35}  & \underline{\textcolor{blue}{63.31}} \\
\bottomrule
\end{tabular}}
\label{Table_smta_ifgsm_nyuv2}
\end{table*}

\begin{table*}[!ht]
\caption{Experimental results of non-stealthy and SMTA$^{2}$ framework by manual and automated solutions of \textbf{IFGSM} attack algorithm on \textbf{Cityscapes} dataset. Values affected by the attack are marked in \textcolor{blue}{blue}, while those not influenced by the attack are highlighted in \textcolor{red}{red}. The performances of the targeted task are underlined. (↑) means higher better and (↓) means lower better.}
\centering
\resizebox{\linewidth}{!}{
\begin{tabular}{c|ccc|ccc|ccc}
\toprule
\begin{tabular}
{@{}c@{}} Method \end{tabular} & \multicolumn{3}{c|}{Non-stealthy} & \multicolumn{3}{c|}{SMTA$^{2}$ Manual} & \multicolumn{3}{c}{SMTA$^{2}$ Auto} \\ \midrule
\begin{tabular}
{@{}c@{}} Tasks \end{tabular} & \begin{tabular}[c]{@{}c@{}}Segment\\ \begin{small}{[}mIoU(↑){]}\end{small}\end{tabular} & \begin{tabular}[c]{@{}c@{}}Part Seg\\ \begin{small}{[}mIoU(↑){]}\end{small}\end{tabular} & \begin{tabular}[c]{@{}c@{}}Disparity\\ \begin{small}{[}aErr(↓){]}\end{small}\end{tabular} &
\begin{tabular}[c]{@{}c@{}}Segment\\ \begin{small}{[}mIoU(↑){]}\end{small}\end{tabular} & \begin{tabular}[c]{@{}c@{}}Part Seg\\ \begin{small}{[}mIoU(↑){]}\end{small}\end{tabular} & \begin{tabular}[c]{@{}c@{}}Disparity\\ \begin{small}{[}aErr(↓){]}\end{small}\end{tabular} & \begin{tabular}[c]{@{}c@{}}Segment\\ \begin{small}{[}mIoU(↑){]}\end{small}\end{tabular} & \begin{tabular}[c]{@{}c@{}}Part Seg\\ \begin{small}{[}mIoU(↑){]}\end{small}\end{tabular} & \begin{tabular}[c]{@{}c@{}}Disparity\\ \begin{small}{[}aErr(↓){]}\end{small}\end{tabular} \\ \midrule
$\epsilon$=0 & 54.20  & 51.82             & 81.51        & 54.20  & 51.82             & 81.51 & 54.20  & 51.82             & 81.51  \\ \midrule
& {\underline{\textcolor{blue}{27.98}}}            & \textcolor{blue}{43.82}                &  \textcolor{blue}{98.72}    & \underline{\textcolor{blue}{25.36}}   & \textcolor{red}{52.48}     & \textcolor{red}{80.20}   & \underline{\textcolor{blue}{25.39}}   & \textcolor{red}{53.31}   & \textcolor{red}{81.17}    \\
$\epsilon$=2/255 &  \textcolor{blue}{41.24}   & {\underline{\textcolor{blue}{30.00}}}  &  \textcolor{blue}{92.13}   & \textcolor{red}{55.33}   &{\underline{\textcolor{blue}{27.49}}}   & 
 \textcolor{red}{79.80}  &   \textcolor{red}{54.57}   &{\underline{\textcolor{blue}{27.21}}}   & 
 \textcolor{red}{76.34} \\  &  \textcolor{blue}{38.47}   &  \textcolor{blue}{42.05}   &  {\underline{\textcolor{blue}{278.11}}}    &  \textcolor{red}{54.59}   &  \textcolor{red}{51.94}   &   {\underline{\textcolor{blue}{366.49}}}  &  \textcolor{red}{56.05}   &  \textcolor{red}{52.29}   &   {\underline{\textcolor{blue}{357.91}}}  \\ 
 \midrule
 & \underline{\textcolor{blue}{20.48}} 
 & \textcolor{blue}{35.01}  &  \textcolor{blue}{119.67}     &  \underline{\textcolor{blue}{17.44}}  &  \textcolor{red}{52.21}  &  \textcolor{red}{80.91}    &  \underline{\textcolor{blue}{17.13}}  &  \textcolor{red}{52.57}  &  \textcolor{red}{81.23} \\  $\epsilon$=4/255 & \textcolor{blue}{33.98} 
  &  \underline{\textcolor{blue}{21.69}} 
  &   \textcolor{blue}{111.19}         & \textcolor{red}{54.58} 
 &  \underline{\textcolor{blue}{18.16}}  
&  \textcolor{red}{81.10}    & \textcolor{red}{54.49} 
 &  \underline{\textcolor{blue}{17.95}}  
 &  \textcolor{red}{80.05}   \\           &  \textcolor{blue}{25.82}  & \textcolor{blue}{26.68}  & \underline{\textcolor{blue}{490.15}}    & \textcolor{red}{57.33}   &   \textcolor{red}{52.51}   &   \underline{\textcolor{blue}{663.81}}  & \textcolor{red}{56.56}   &   \textcolor{red}{51.89}    &  \underline{\textcolor{blue}{648.99}}   \\
\midrule
&  \underline{\textcolor{blue}{11.31}}  &  \textcolor{blue}{19.24}  &  \textcolor{blue}{164.19}  & 
 \underline{\textcolor{blue}{10.31}}  &   \textcolor{red}{51.87}   &  \textcolor{red}{81.14}  &  \underline{\textcolor{blue}{9.93}}  &   \textcolor{red}{51.99}   &  \textcolor{red}{81.85} \\ $\epsilon$=8/255   & \textcolor{blue}{25.77} 
&  \underline{\textcolor{blue}{14.65}}   &  \textcolor{blue}{153.10}  &    \textcolor{red}{54.87} 
&  \underline{\textcolor{blue}{12.54}}  & \textcolor{red}{81.41}   &   \textcolor{red}{54.54}  
&  \underline{\textcolor{blue}{12.91}}  &   \textcolor{red}{79.45}  \\   & \textcolor{blue}{16.05}  &   \textcolor{blue}{14.18}   &   \underline{\textcolor{blue}{824.75}}     &   \textcolor{red}{55.52}  &  \textcolor{red}{52.01}  & \underline{\textcolor{blue}{1016.41}}  &   \textcolor{red}{54.75}  &  \textcolor{red}{51.97}  & \underline{\textcolor{blue}{1029.93}} \\
\bottomrule
\end{tabular}}
\label{Table_smta_ifgsm_city}
\end{table*}

\begin{table*}[!ht]
\caption{Experimental results of non-stealthy and SMTA$^{2}$ framework \textbf{with adversarial training} by manual and automated solutions of \textbf{IFGSM} attack algorithm on \textbf{NYUv2} dataset. Values affected by the attack are marked in \textcolor{blue}{blue}, while those not influenced by the attack are highlighted in \textcolor{red}{red}. The performances of the targeted task are underlined. (↑) means higher better and (↓) means lower better.}
\centering
\resizebox{\linewidth}{!}{
\begin{tabular}{c|ccc|ccc|ccc}
\toprule
\begin{tabular}
{@{}c@{}} Method \end{tabular} & \multicolumn{3}{c|}{Non-stealthy} & \multicolumn{3}{c|}{SMTA$^{2}$ Manual} & \multicolumn{3}{c}{SMTA$^{2}$ Auto} \\ \midrule
\begin{tabular}
{@{}c@{}} Tasks \end{tabular} & \begin{tabular}[c]{@{}c@{}}Segment\\ \begin{small}{[}mIoU(↑){]}\end{small}\end{tabular} & \begin{tabular}[c]{@{}c@{}}Depth\\ \begin{small}{[}aErr(↓){]}\end{small}\end{tabular} & \begin{tabular}[c]{@{}c@{}}Normal\\ \begin{small}{[}mDist(↓){]}\end{small}\end{tabular} & \begin{tabular}[c]{@{}c@{}}Segment\\ \begin{small}{[}mIoU(↑){]}\end{small}\end{tabular} & \begin{tabular}[c]{@{}c@{}}Depth\\ \begin{small}{[}aErr(↓){]}\end{small}\end{tabular} & \begin{tabular}[c]{@{}c@{}}Normal\\ \begin{small}{[}mDist(↓){]}\end{small}\end{tabular} & \begin{tabular}[c]{@{}c@{}}Segment\\ \begin{small}{[}mIoU(↑){]}\end{small}\end{tabular} & \begin{tabular}[c]{@{}c@{}}Depth\\ \begin{small}{[}aErr(↓){]}\end{small}\end{tabular} & \begin{tabular}[c]{@{}c@{}}Normal\\ \begin{small}{[}mDist(↓){]}\end{small}\end{tabular} \\ \midrule
$\epsilon$=0 & 46.56  & 40.57                           & 23.41        & 46.56  & 40.57                           & 23.41   & 46.56  & 40.57                           & 23.41   \\ \midrule
& \underline{\textcolor{blue}{29.46}}                       & \textcolor{blue}{58.56}               &  \textcolor{blue}{29.99} & \underline{\textcolor{blue}{29.99}}   &  \textcolor{red}{39.76}  & \textcolor{red}{22.30}  & \underline{\textcolor{blue}{29.81}}   & \textcolor{red}{39.84} &  \textcolor{red}{22.55}      \\
$\epsilon$=2/255  &  \textcolor{blue}{35.53}   & \underline{\textcolor{blue}{57.00}}  &  \textcolor{blue}{28.04}   & \textcolor{red}{47.38}   & \underline{\textcolor{blue}{60.97}} & 
\textcolor{red}{23.28}   &  \textcolor{red}{47.33}   & \underline{\textcolor{blue}{61.21}} & \textcolor{red}{22.50}
\\  &  \textcolor{blue}{35.45}   &  \textcolor{blue}{52.27}   &  \underline{\textcolor{blue}{29.50}}    &  \textcolor{red}{46.80} & \textcolor{red}{40.15}   & \underline{\textcolor{blue}{31.01}}   &  \textcolor{red}{46.67}   &  \textcolor{red}{40.07}   & \underline{\textcolor{blue}{30.65}}  \\ 
 \midrule
 & \underline{\textcolor{blue}{30.14}} 
 & \textcolor{blue}{53.43}  &  \textcolor{blue}{28.37}    &  \underline{\textcolor{blue}{30.81}}  &  \textcolor{red}{39.65}  &  \textcolor{red}{23.30}   &  \underline{\textcolor{blue}{30.33}}  &  \textcolor{red}{39.24}  &  \textcolor{red}{23.32}   \\ $\epsilon$=4/255 & \textcolor{blue}{34.58} 
  &  \underline{\textcolor{blue}{63.36}} 
  &   \textcolor{blue}{28.46}     & \textcolor{red}{46.83} 
 &  \underline{\textcolor{blue}{60.59}}  
&  \textcolor{red}{23.04}     & \textcolor{red}{46.84} 
 &  \underline{\textcolor{blue}{62.32}}  
 &  \textcolor{red}{23.04}      \\           &  \textcolor{blue}{30.41}  & \textcolor{blue}{59.41}  & \underline{\textcolor{blue}{32.59}}    & \textcolor{red}{47.54}   &   \textcolor{red}{39.22}   &   \underline{\textcolor{blue}{31.42}}  & \textcolor{red}{48.02} &   \textcolor{red}{39.45}    &   \underline{\textcolor{blue}{30.55}}   \\
\midrule
&  \underline{\textcolor{blue}{23.33}}  &  \textcolor{blue}{61.37} &  \textcolor{blue}{30.97}  & \underline{\textcolor{blue}{25.02}}
 & \textcolor{red}{37.80}   & \textcolor{red}{22.75}   & \underline{\textcolor{blue}{24.58}}  & \textcolor{red}{39.02}   & \textcolor{red}{23.26} 
 \\ $\epsilon$=8/255   & \textcolor{blue}{28.68}
&  \underline{\textcolor{blue}{76.32}}   &  \textcolor{blue}{31.15}   & \textcolor{red}{47.38}   
& \underline{\textcolor{blue}{39.60}}  & \textcolor{red}{23.37}   &   \textcolor{red}{47.27}
& \underline{\textcolor{blue}{38.82}}  &  \textcolor{red}{22.59} \\   & \textcolor{blue}{28.90} &   \textcolor{blue}{61.44}  &  \underline{\textcolor{blue}{35.85}}   &  \textcolor{red}{47.00}  & \textcolor{red}{39.95}  & \underline{\textcolor{blue}{35.27}}   & \textcolor{red}{46.63}   & \textcolor{red}{37.70}  & \underline{\textcolor{blue}{34.77}} \\
\bottomrule
\end{tabular}}
\label{Table_smta_at_ifgsm_nyuv2}
\end{table*}

\begin{table*}[!ht]
\caption{Experimental results of non-stealthy and SMTA$^{2}$ framework \textbf{with adversarial training} by manual and automated solutions of \textbf{IFGSM} attack algorithm on \textbf{Cityscapes} dataset. Values affected by the attack are marked in \textcolor{blue}{blue}, while those not influenced by the attack are highlighted in \textcolor{red}{red}. The performances of the targeted task are underlined. (↑) means higher better and (↓) means lower better.}
\centering
\resizebox{\linewidth}{!}{
\begin{tabular}{c|ccc|ccc|ccc}
\toprule
\begin{tabular}
{@{}c@{}} Method \end{tabular} & \multicolumn{3}{c|}{Non-stealthy} & \multicolumn{3}{c|}{SMTA$^{2}$ Manual} & \multicolumn{3}{c}{SMTA$^{2}$ Auto} \\ \midrule
\begin{tabular}
{@{}c@{}} Tasks \end{tabular} & \begin{tabular}[c]{@{}c@{}}Segment\\ \begin{small}{[}mIoU(↑){]}\end{small}\end{tabular} & \begin{tabular}[c]{@{}c@{}}Part Seg\\ \begin{small}{[}mIoU(↑){]}\end{small}\end{tabular} & \begin{tabular}[c]{@{}c@{}}Disparity\\ \begin{small}{[}aErr(↓){]}\end{small}\end{tabular} &
\begin{tabular}[c]{@{}c@{}}Segment\\ \begin{small}{[}mIoU(↑){]}\end{small}\end{tabular} & \begin{tabular}[c]{@{}c@{}}Part Seg\\ \begin{small}{[}mIoU(↑){]}\end{small}\end{tabular} & \begin{tabular}[c]{@{}c@{}}Disparity\\ \begin{small}{[}aErr(↓){]}\end{small}\end{tabular} & \begin{tabular}[c]{@{}c@{}}Segment\\ \begin{small}{[}mIoU(↑){]}\end{small}\end{tabular} & \begin{tabular}[c]{@{}c@{}}Part Seg\\ \begin{small}{[}mIoU(↑){]}\end{small}\end{tabular} & \begin{tabular}[c]{@{}c@{}}Disparity\\ \begin{small}{[}aErr(↓){]}\end{small}\end{tabular} \\ \midrule
$\epsilon$=0 & 54.20  & 51.82             & 81.51        & 54.20  & 51.82             & 81.51 & 54.20  & 51.82             & 81.51  \\ \midrule
& \underline{\textcolor{blue}{33.21}}            & \textcolor{blue}{38.61}                &  \textcolor{blue}{101.67}    & \underline{\textcolor{blue}{33.95}}   & \textcolor{red}{52.42}     & \textcolor{red}{80.09}   & \underline{\textcolor{blue}{33.84}}   & \textcolor{red}{52.43}   &  \textcolor{red}{80.99}    \\
$\epsilon$=2/255 &  \textcolor{blue}{40.59}  & \underline{\textcolor{blue}{41.09}}  &  \textcolor{blue}{94.69}   & \textcolor{red}{55.06}   & \underline{\textcolor{blue}{37.38}} & \textcolor{red}{80.99}
&  \textcolor{red}{54.14}    & \underline{\textcolor{blue}{37.13}}  & \textcolor{red}{80.62}
\\  &  \textcolor{blue}{33.69}   &  \textcolor{blue}{37.34}   &  \underline{\textcolor{blue}{103.34}}    &  \textcolor{red}{56.83}  & \textcolor{red}{52.32}    & \underline{\textcolor{blue}{105.14}}    & \textcolor{red}{54.98}   &  \textcolor{red}{52.34}   &  \underline{\textcolor{blue}{100.48}}   \\ 
 \midrule
 & \underline{\textcolor{blue}{30.98}} 
 & \textcolor{blue}{37.93}  &  \textcolor{blue}{102.75}     &  \underline{\textcolor{blue}{30.18}}  &  \textcolor{red}{52.21}  &  \textcolor{red}{81.03}    &  \underline{\textcolor{blue}{31.08}}  &  \textcolor{red}{52.37}  &  \textcolor{red}{80.92} \\  $\epsilon$=4/255 & \textcolor{blue}{34.17} 
  &  \underline{\textcolor{blue}{34.78}} 
  &   \textcolor{blue}{101.93}         & \textcolor{red}{54.39} 
 &  \underline{\textcolor{blue}{31.06}}  
&  \textcolor{red}{81.43}    & \textcolor{red}{54.49} 
 &  \underline{\textcolor{blue}{31.25}}  
 &  \textcolor{red}{81.77}   \\           &  \textcolor{blue}{34.41}  & \textcolor{blue}{38.24}  & \underline{\textcolor{blue}{112.85}}    & \textcolor{red}{55.23}   &   \textcolor{red}{52.02}   &   \underline{\textcolor{blue}{115.09}}  & \textcolor{red}{55.26}   &   \textcolor{red}{52.98}    &  \underline{\textcolor{blue}{114.03}}   \\
\midrule
&  \underline{\textcolor{blue}{30.28}}  &  \textcolor{blue}{48.46}  &  \textcolor{blue}{92.85}  & 
\underline{\textcolor{blue}{34.26}}  & \textcolor{red}{52.11}    & \textcolor{red}{78.23}   & \underline{\textcolor{blue}{34.26}}  &   \textcolor{red}{53.45}   &  \textcolor{red}{80.87} \\ $\epsilon$=8/255   & \textcolor{blue}{43.37} 
&  \underline{\textcolor{blue}{31.76}}   & \textcolor{blue}{90.25}   &    \textcolor{red}{54.30} 
&  \underline{\textcolor{blue}{35.71}}  & \textcolor{red}{80.71}   &   \textcolor{red}{55.38}  
&  \underline{\textcolor{blue}{31.76}}  &   \textcolor{red}{81.29}  \\   & \textcolor{blue}{40.26}  &   \textcolor{blue}{38.72}   &   \underline{\textcolor{blue}{122.16}}     &   \textcolor{red}{54.27}  &  \textcolor{red}{52.43}  & \underline{\textcolor{blue}{120.33}}  &   \textcolor{red}{54.28}  &  \textcolor{red}{52.73}  & \underline{\textcolor{blue}{122.11}} \\
\bottomrule
\end{tabular}}
\label{Table_smta_at_ifgsm_city}
\end{table*}

\begin{figure*}[!ht]
\centering
\includegraphics[width=\linewidth]{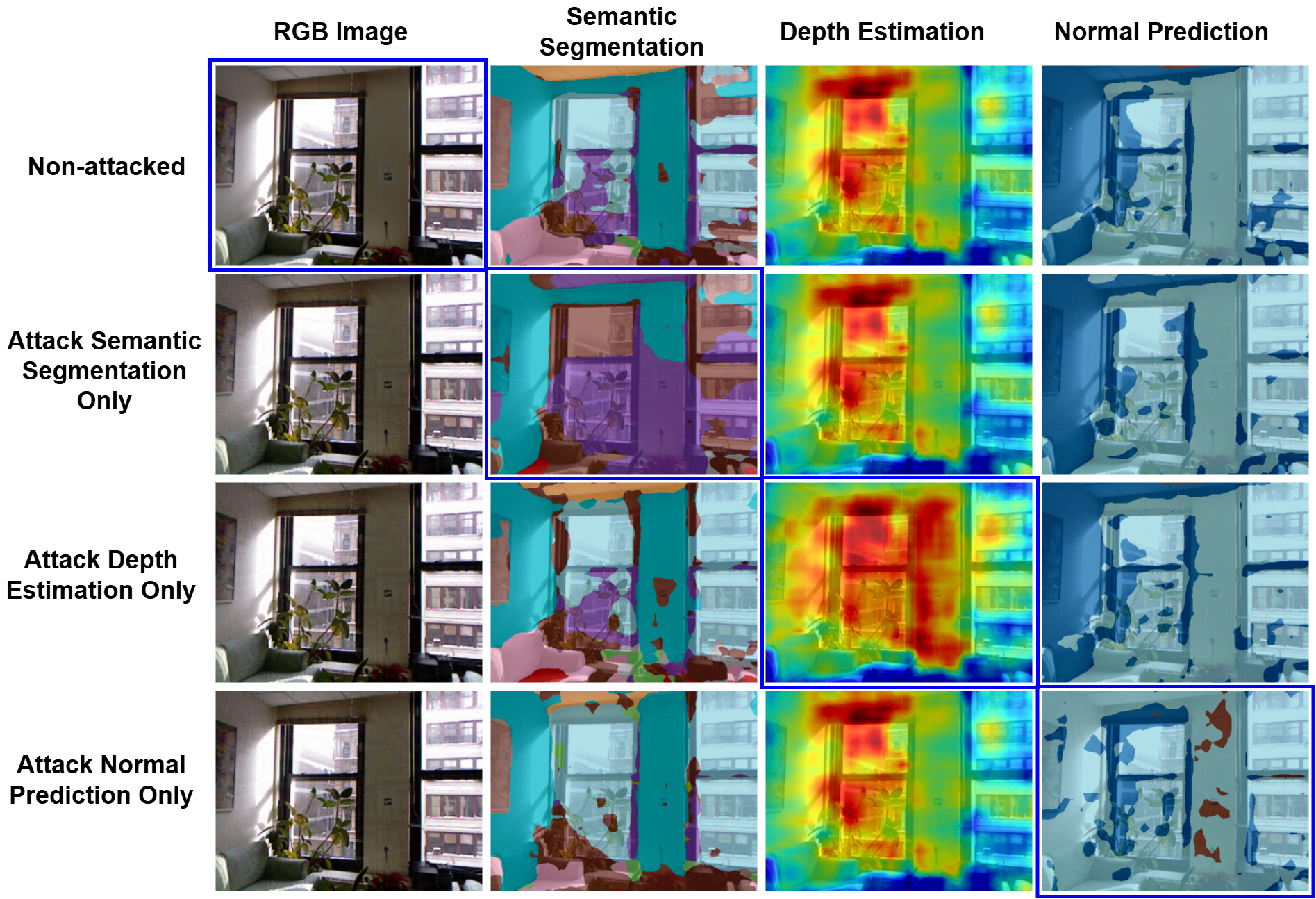} 
\caption{Example output images of non-attacked output and SMTA$^{2}$ model attacking three tasks respectively on the NYUv2 dataset 
($\epsilon$=2/255). Each row shows the targeted task being attacked, and each column shows the performance of targeted and non-targeted tasks. The images enclosed in boxes include the original input and the target outputs of the tasks being selectively attacked. While the targeted task is well-attacked, the SMTA$^{2}$ framework preserves the performance of non-targeted tasks.}
\label{Atk_nyuv2_1}
\end{figure*}

\begin{figure*}[!ht]
\centering
\includegraphics[width=\linewidth]{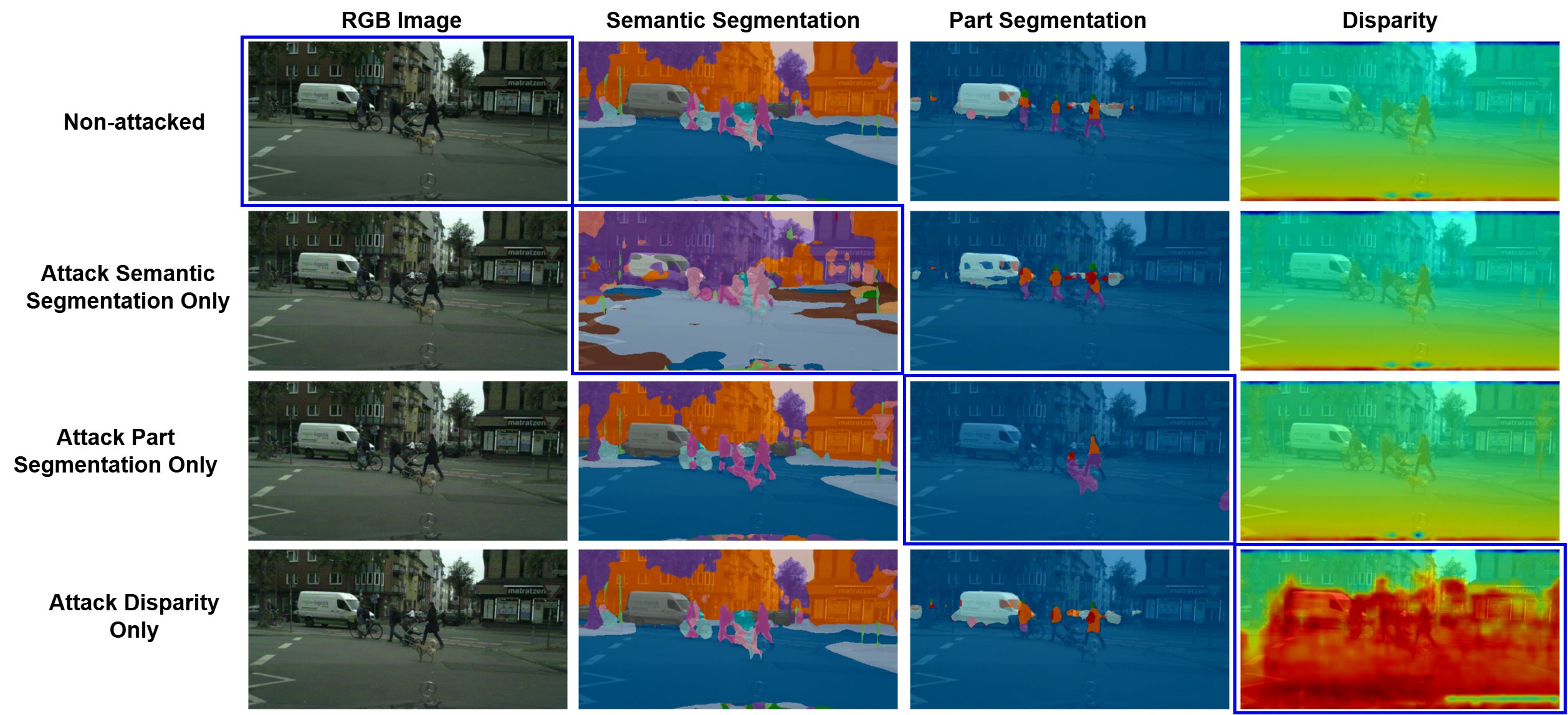} 
\caption{{Example output images of non-attacked and SMTA$^{2}$ models attacking three tasks respectively on the Cityscapes dataset ($\epsilon$=2/255). Each row shows the targeted task being attacked, and each column shows the performance of targeted and non-targeted tasks. The images enclosed in boxes include the original input and the target outputs of the tasks being selectively attacked. While the targeted task is well-attacked, the SMTA$^{2}$ framework preserves the original performance of non-targeted tasks.}}
\label{Atk_city_1}
\end{figure*}

\begin{figure}[!ht]
\centering
\includegraphics[width=\linewidth]{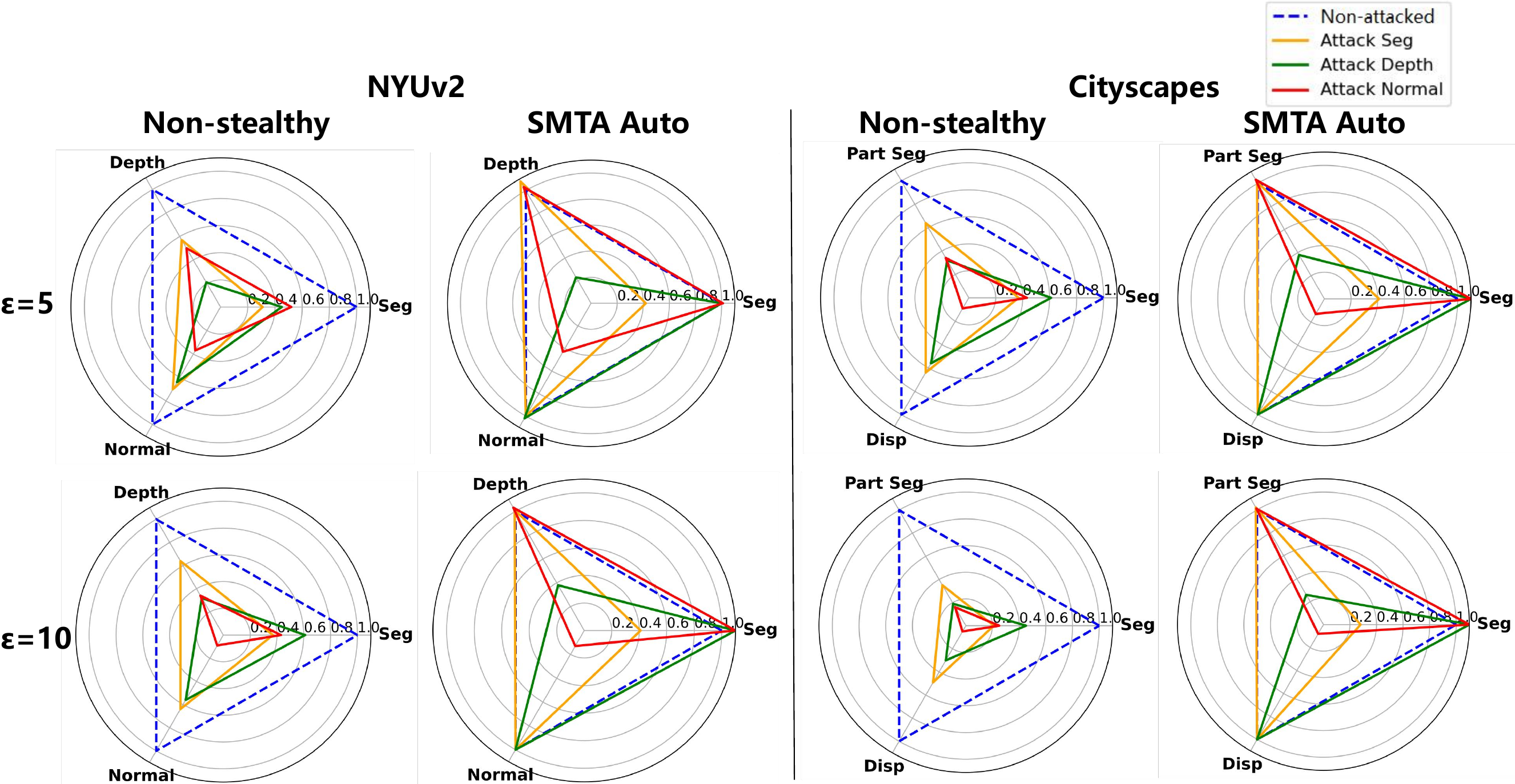} 
\caption{Radar images of undefended PGD $L_2$ attack effectiveness of non-stealthy and SMTA$^{2}$ Automated methods of NYUv2 and Cityscapes datasets.}
\label{radar_pgdl2_undefended}
\end{figure}

Figure~\ref{radar_pgdl2_undefended} illustrates the visualized attack effectiveness of non-stealthy and automated SMTA$^{2}$ methods under PGD $L_2$ attack, respectively. These radar charts depict the ratio between post-attack and pre-attack performance, serving as a measure of attack effectiveness.

\begin{figure}[!ht]
\centering
\includegraphics[width=1\linewidth]{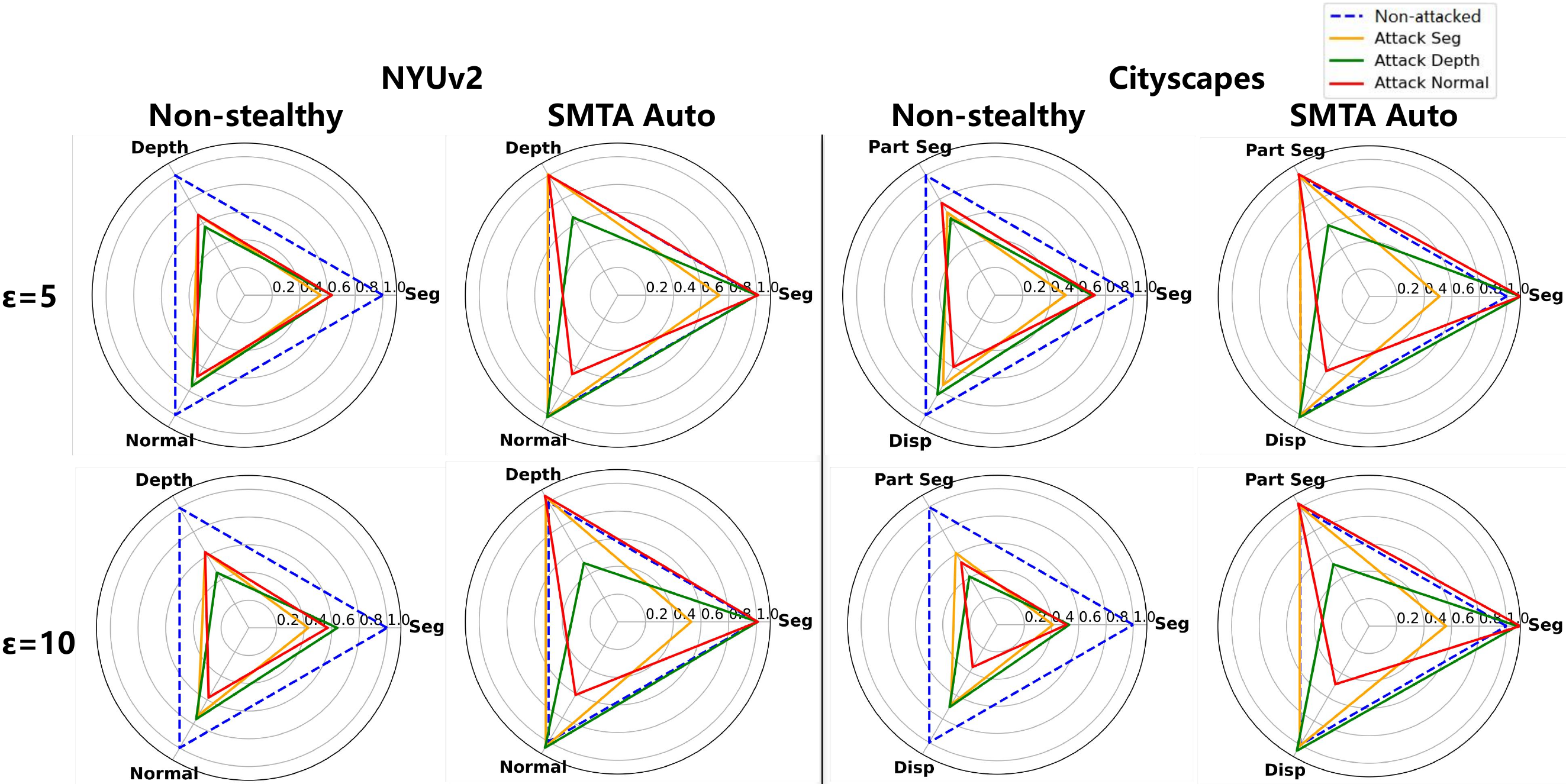} 
\caption{Radar images of \textbf{Adversarially Trained PGD $L_2$}  attack effectiveness of non-stealthy and SMTA$^{2}$ Automated methods of NYUv2 and Cityscapes datasets.}
\label{radar_pgdl2_at}
\end{figure}

Figure~\ref{radar_pgdl2_at} presents the visualized attack effectiveness of both non-stealthy and automated SMTA$^{2}$ methods under PGD $L_2$ attacks, based on an Adversarially Trained (AT) model. Similar to the results in Figure~\ref{radar_pgdl2_undefended}, the non-stealthy attack demonstrates that, even when only a single task is targeted, other tasks in a conventional MTA setting are also adversely affected. In contrast, the proposed SMTA$^{2}$ method effectively preserves the original performance of all non-targeted tasks, even under adversarial perturbations. Furthermore, when comparing the AT model (Figure~\ref{radar_pgdl2_at}) with the undefended model, it is evident that adversarial training reduces the overall impact of attacks to some extent. Nevertheless, SMTA$^{2}$ sustainably to achieve effective degradation on the targeted task while simultaneously safeguarding the performance of non-targeted tasks.

\subsection{Results Compared with Single-task Attacks}

The attack effectiveness is compared between the single-task attack model and our proposed SMTA$^{2}$ framework using the automated solution with PGD and IFGSM attack. In the PGD algorithm, both $L_\infty$ and $L_2$ norms are implemented. For IFGSM and PGD $L_\infty$ attacks, {the perturbation $\epsilon$ is set to} 4/255, while for the PGD $L_2$ attack, {$\epsilon$=$10$ is implemented} for a fair comparison. ``$\epsilon$=0'' represents the model remains original training state without any attack.

In the single-task attack model, a specific task is targeted without affecting other tasks. As shown in Table \ref{Table_sta}, for the NYUv2 dataset, the single-task attack model is generally more effective. However, for the Cityscapes dataset, the SMTA$^{2}$ framework with automated weight searching outperforms the single-task model across all attack algorithms. Since both the PGD $L_\infty$ and IFGSM algorithms utilize the $L_\infty$ norm, their attack performances are quite similar. Overall, the SMTA$^{2}$ framework achieves comparable attack effectiveness to the single-task approach while ensuring that non-targeted tasks maintain their performance, and in some cases, it even outperforms the single-task attack.

\begin{table*}[!ht]
\caption{Results of single-task attack models on NYUv2 and Cityscapes datasets compared with the automated weighting method of SMTA$^{2}$ of selected perturbation. The results of SMTA$^{2}$ based on the automated weighting method are subscripted. The better attack result is highlighted in \textbf{bold}.}
\centering
\resizebox{\linewidth}{!}{
\begin{tabular}{
c|c|ccc|ccc}
\toprule
\multicolumn{2}{c|}{\begin{tabular}{@{}c@{}} Dataset \end{tabular}} & \multicolumn{3}{c|}{NYUv2} &  \multicolumn{3}{c}{Cityscapes} \\ \midrule
Method & Perturbation & \begin{tabular}[c]{@{}c@{}}Segment\\ \begin{small}{[}mIoU(↑){]}\end{small}\end{tabular} & \begin{tabular}[c]{@{}c@{}}Depth\\ \begin{small}{[}aErr(↓){]}\end{small}\end{tabular} & \begin{tabular}[c]{@{}c@{}}Normal\\ \begin{small}{[}mDist(↓){]}\end{small}\end{tabular} & \begin{tabular}[c]{@{}c@{}}Segment\\ \begin{small}{[}mIoU(↑){]}\end{small}\end{tabular} & \begin{tabular}[c]{@{}c@{}}Part Seg\\ \begin{small}{[}mIoU(↑){]}\end{small}\end{tabular} & \begin{tabular}[c]{@{}c@{}}Disparity\\ \begin{small}{[}aErr(↓){]}\end{small}\end{tabular} \\ \midrule
& $\epsilon$=0      & \textbf{43.26} \begin{scriptsize}{(46.56)}
\end{scriptsize}         & \textbf{54.57} \begin{scriptsize}{(40.57)}
\end{scriptsize}     & 22.35 \begin{scriptsize}\textbf{{(23.41)}}
\end{scriptsize} & 57.81 \begin{scriptsize}\textbf{{(54.20)}}
\end{scriptsize}         & 54.20 \begin{scriptsize}\textbf{{(51.82)}}
\end{scriptsize}     & \textbf{83.23} \begin{scriptsize}{(81.51)}
\end{scriptsize}    \\ \midrule
IFGSM                                 & $\epsilon$=4/255    & \textbf{15.81} \begin{scriptsize}{(18.33)}
\end{scriptsize}      & \textbf{188.75} \begin{scriptsize}{(139.30)}
\end{scriptsize}      & \textbf{58.13} \begin{scriptsize}{(52.55)}
\end{scriptsize}    & 20.43 \begin{scriptsize}\textbf{{(17.13)}}
\end{scriptsize}      & 23.40 \begin{scriptsize}\textbf{{(17.95)}}
\end{scriptsize}      & 500.03 \begin{scriptsize}\textbf{{(648.99)}}
\end{scriptsize}  \\
PGD $L_\infty$                             & $\epsilon$=4/255                                  & \textbf{15.81} \begin{scriptsize}{(18.33)}
\end{scriptsize}     & \textbf{188.74} \begin{scriptsize}{(168.69)}
\end{scriptsize}               & \textbf{58.17} \begin{scriptsize}{(53.87)}
\end{scriptsize}    & 20.44 \begin{scriptsize}\textbf{{(17.08)}}
\end{scriptsize}     & 23.33 \begin{scriptsize}\textbf{{(18.19)}}
\end{scriptsize}               & 499.70 \begin{scriptsize}\textbf{{(644.12)}}
\end{scriptsize}   \\
PGD $L_2$                                & $\epsilon$=10                                     & \textbf{12.59} \begin{scriptsize}{(12.84)}
\end{scriptsize}      & 249.72 \begin{scriptsize}\textbf{{(251.15)}}
\end{scriptsize}      & 67.23 \begin{scriptsize}\textbf{{(68.21)}}
\end{scriptsize}    & 21.26 \begin{scriptsize}\textbf{{(14.25)}}
\end{scriptsize}      & 21.53 \begin{scriptsize}\textbf{{(13.49)}}
\end{scriptsize}      & 601.42 \begin{scriptsize}\textbf{{(1026.31)}}
\end{scriptsize}  \\ \bottomrule 
\end{tabular}}
\label{Table_sta}
\end{table*}

%
%
\clearpage
\bibliographystyle{splncs04}
\bibliography{references}